\documentclass[aps,prd,twocolumn,superscriptaddress]{revtex4}
\usepackage{epsfig,epsf}
\usepackage{amsmath}
\usepackage{amsthm}
\usepackage{amsfonts}
\usepackage{amssymb}
\usepackage{dsfont}
\usepackage{multirow}
\usepackage{appendix}
\usepackage{slashed}
\usepackage[active]{srcltx}
\usepackage{psfrag}

\begin{document}

\title{Scalar molecules $\eta _{b}B_{c}^{-}$ and $\eta _{c}B_{c}^{+} $ with
asymmetric quark contents}
\date{\today}
\author{S.~S.~Agaev}
\affiliation{Institute for Physical Problems, Baku State University, Az--1148 Baku,
Azerbaijan}
\author{K.~Azizi}
\thanks{Corresponding Author}
\affiliation{Department of Physics, University of Tehran, North Karegar Avenue, Tehran
14395-547, Iran}
\affiliation{Department of Physics, Dogus University, Dudullu-\"{U}mraniye, 34775
Istanbul, T\"{u}rkiye}
\author{H.~Sundu}
\affiliation{Department of Physics Engineering, Istanbul Medeniyet University, 34700
Istanbul, T\"{u}rkiye}

\begin{abstract}
The hadronic scalar molecules $\mathcal{M}_{b}$ and $\mathcal{M}_{c}$ with
asymmetric quark contents $bb \overline{b}\overline{c}$ and $cc \overline{c}
\overline{b}$ are explored by means of the QCD sum rule method. Their masses
and current couplings are calculated using the two-point sum rule approach.
The obtained results show that they are strong-interaction unstable
particles and transform to ordinary mesons' pairs. The molecule $\mathcal{M}
_{b}$ dissociates through the process $\mathcal{M}_{\mathrm{b}}\to \eta
_{b}B_{c}^{-}$. The decays $\mathcal{M}_{\mathrm{c}}\rightarrow \eta
_{c}B_{c}^{+}$ and $J/\psi B_{c}^{\ast +}$ are dominant modes for the
molecule $\mathcal{M}_{c}$. The full decay widths of the molecules $\mathcal{%
\ \ M}_{b}$ and $\mathcal{M}_{c}$ are estimated using these decay channels,
as well as ones generated by the annihilation of $b\overline{b}$ and $c
\overline{c}$ quarks in $\mathcal{M}_{b}$ and $\mathcal{M}_{c}$,
respectively. The QCD three-point sum rule method is employed to find
partial widths all of these channels. This approach is required to evaluate
the strong couplings at the molecule-meson-meson vertices under
consideration. The mass $m=(15728 \pm 90)~\mathrm{MeV}$ and width $\Gamma[
\mathcal{M}_b] =(93 \pm 17)~ \mathrm{MeV}$ of the molecule $\mathcal{M}_{b}$
, and $\widetilde{m}=(9712 \pm 72)~\mathrm{MeV}$ and $\Gamma[\mathcal{M}_c]
=(70 \pm 10)~ \mathrm{MeV}$ in the case of $\mathcal{M}_{c} $ offer valuable
guidance for experimental searches at existing facilities.
\end{abstract}

\maketitle

%%%%%%%%%%%%%%%%%%%%%%%%%%%%%%%%%%%%%%%%%%%%%%%%%%%%%%%%%%%%%%%%%

\section{Introduction}

\label{sec:Intro}
%%%%%%%%%%%%%%%%%%%%%%%%%%%%%%%%%%%%%%%%%%%%%%%%%%%%%%%%%%%

Hadronic four-quark exotic molecular states are already on agenda of high
energy physics. Such structures may appear in experiments as a bound and/or
resonant states of a pair of ordinary mesons. These molecules are composed
of the color-singlet quark-antiquarks, and have internal organizations
alternative to those of diquark-antidiquarks: In a diquark-antidiquark
picture four-quark mesons are built of colored diquarks and antidiquarks.

Theoretical investigations of hadronic molecules have a rather long history.
Thus, existence of the hadronic molecules $c\overline{q}\overline{c}q$ were
supposed in Ref. \cite{Bander:1975fb} in light of numerous vector states $J^{%
\mathrm{PC}}=1^{--}$ observed in $e^{+}e^{-}$ annihilation. Analogous ideas
were shared by the authors of the publications \cite%
{Voloshin:1976ap,DeRujula:1976zlg}, in which they suggested that four-quark
mesons may emerge as bound-resonant states of the $D$ mesons, interacting
via conventional light meson exchange mechanism.

The concept of hadronic molecules was later elaborated and advanced in
numerous investigations \cite%
{Tornqvist:1991lks,Ding:2008mp,Zhang:2009vs,Sun:2012sy,Chen:2015ata,Karliner:2015ina, Liu:2016kqx,Chen:2017vai,Sun:2018zqs,Albuquerque:2012rq,PavonValderrama:2019ixb,Molina:2020hde,Xu:2020evn, Xin:2021wcr,Agaev:2022duz,Agaev:2023eyk,Braaten:2023vgs,Wu:2023rrp,Liang:2023jxh,Wang:2025zss,Braaten:2024tbm}%
, in which the authors explored the binding mechanisms of such states,
computed their masses, analyzed processes where these particles might be
discovered. Needless to say that all available models and methods were
applied in these studies to reach reliable conclusions about properties of
hadronic molecules.

Another interesting branch of investigations embraces molecules containing
only heavy $c$ and $b$ quarks. They may consist of only $c$ ($b$) quarks, or
may be composed of equal number of these quarks. These molecules are hidden
charm, bottom, or charm-bottom particles. The molecules of the first type
were examined in Refs.\ \cite%
{Agaev:2023ruu,Agaev:2023rpj,Yalikun:2025ssz,Liu:2024pio}. Activity of
researches in this field was inspired mainly by observation of new four $X$
structures reported by LHCb-ATLAS-CMS collaborations \cite%
{LHCb:2020bwg,ATLAS:2023bft,CMS:2023owd}. These structures are presumably
scalar resonances made of $cc\overline{c}\overline{c}$ quarks. It turns out
that some of them may be interpreted as hadronic molecules.

Relevant problems were also addressed in our works \cite%
{Agaev:2023ruu,Agaev:2023rpj}, in which we considered fully heavy molecules $%
\eta _{c}\eta _{c}$, $\chi _{c0}\chi _{c0}$, and $\chi _{c1}\chi _{c1}$ and
computed their masses and decay widths. Our aim was to compare obtained
results with measured parameters of different $X$ structures. We argued that
the molecule $\eta _{c}\eta _{c}$ can be considered as a real candidate to
the resonance $X(6200)$, whereas the structure $\chi _{c0}\chi _{c0}$ may be
interpreted as $X(6900)$ or one of its components in combination with a
scalar diquark-antidiquark state. The mass and width of the molecule $\chi
_{c1}\chi _{c1}$ is comparable with those of the structure $X(7300)$, but
preferable model for this structure is an admixture of $\chi _{c1}\chi _{c1}$
with sizeable excited diquark-antidiquark component.

There are also various publications devoted to analysis of the molecules
with mixed contents \cite%
{Liu:2024pio,Liu:2023gla,Wang:2023bek,Agaev:2025wdj,Agaev:2025fwm,Agaev:2025nkw}%
. The molecules $B_{c}^{(\ast )+}B_{c}^{(\ast )-}$ were considered in Ref.
\cite{Liu:2023gla} in the context of the coupled-channel unitary approach.
The parameters of the scalar $B_{c}^{+}B_{c}^{-}$, axial-vector $%
(B_{c}^{\ast +}B_{c}^{-}+B_{c}^{+}B_{c}^{\ast -})/2$ and tensor $B_{c}^{\ast
+}B_{c}^{\ast -}$ mesons were calculated in our articles \cite%
{Agaev:2025wdj,Agaev:2025fwm,Agaev:2025nkw}. There, we applied QCD sum rule
(SR) method to evaluate masses and full decay widths of these molecules.

Exotic mesons with the asymmetric quark structures $bb\overline{b}\overline{c%
}$ and $cc\overline{c}\overline{b}$ also attracted interest of researches.
Properties of such diquark-antidiquarks with different spin-parities were
investigated in various works (see, the publications \cite%
{Galkin:2023wox,Agaev:2024uza} and references therein). The hadronic
molecules with the same features were considered in Ref.\ \cite{Liu:2024pio}.

In present work, we explore the scalar heavy hadronic molecules $\mathcal{M}%
_{\mathrm{b}}=\eta _{b}B_{c}^{-}$ and $\mathcal{M}_{\mathrm{c}}=\eta
_{c}B_{c}^{+}$ by computing their masses and full decay widths. They have
quark contents $bb\overline{b}\overline{c}$ and $cc\overline{c}\overline{b}$%
, and evidently are molecular analogues of the asymmetric tetraquarks $T_{%
\mathrm{b}}$ and $T_{\mathrm{c}}$ \cite{Agaev:2024uza}. Investigations are
carried out in the framework of the two-point QCD SR method \cite%
{Shifman:1978bx,Shifman:1978by,Albuquerque:2018jkn,Agaev:2020zad,Wang:2025sic}. Results obtained for the masses of these
structures imply that they are strong-interaction unstable particles and
convert to a pair of ordinary mesons. The molecule $\mathcal{M}_{\mathrm{b}}$
dissociates to its components $\mathcal{M}_{\mathrm{b}}\rightarrow \eta
_{b}B_{c}^{-}$. Apart from this dominant channel, due to annihilation of $b%
\overline{b}$ quarks, $\mathcal{M}_{\mathrm{b}}$ can transform to pairs of
pseudoscalar $B^{-}\overline{D}^{0}$, $\overline{B}^{0}D^{-}$, $\overline{B}%
_{s}^{0}D_{s}^{-}$, and vector $B^{\ast -}\overline{D}^{\ast 0}$, $\overline{%
B}^{\ast 0}D^{\ast -}$, $\overline{B}_{s}^{\ast 0}D_{s}^{\ast -}$ mesons.
Importance of this mechanism was emphasized in Refs.\ \cite%
{Becchi:2020mjz,Becchi:2020uvq,Agaev:2023ara} and applied there to
diquark-antidiquark mesons.

Dominant channels of the state $\mathcal{M}_{\mathrm{c}}$ are the decays $%
\mathcal{M}_{\mathrm{c}}\rightarrow \eta _{c}B_{c}^{+}$, $J/\psi B_{c}^{\ast
+}$, as well as processes $\mathcal{M}_{\mathrm{c}}\rightarrow B^{+}D^{0}$, $%
B^{0}D^{+}$, $B_{s}^{0}D_{s}^{+}$, $B^{\ast +}D^{\ast 0}$, $B^{\ast
0}D^{\ast +}$, and $B_{s}^{\ast 0}D_{s}^{\ast +}$. The last six modes are
generated because of the $c\overline{c}$ annihilation in\ $\mathcal{M}_{%
\mathrm{c}}$.

The widths of the decay channels depend on numerous input parameters of the
molecules $\mathcal{M}_{\mathrm{b}}$ and $\mathcal{M}_{\mathrm{c}}$, and
those of final-state mesons. The masses and couplings of $\mathcal{M}_{%
\mathrm{b}}$ and $\mathcal{M}_{\mathrm{c}}$ are object of the present
studies. The parameters of the conventional mesons are known from
experimental measurements or were found using different theoretical methods.
Decisive quantities which should be determined are the strong couplings at,
for instance, the vertices $\mathcal{M}_{\mathrm{b}}\eta _{b}B_{c}^{-}$, $%
\mathcal{M}_{\mathrm{c}}\eta _{c}B_{c}^{+}$ and $\mathcal{M}_{\mathrm{c}%
}J/\psi B_{c}^{\ast +}$. They describe the strong interaction of the
molecule with ordinary final-state mesons and can be estimated by means of
the QCD three-point sum rule method that allows one to evaluate relevant
form factors.

This paper is organized in the following way: In Sec.\ \ref{sec:Mass}, we
compute the masses and current couplings of the scalar molecules $\mathcal{M}%
_{\mathrm{b}}$ and $\mathcal{M}_{\mathrm{c}}$. The width of the molecule $%
\mathcal{M}_{\mathrm{b}}$ is calculated in Sec.\ \ref{sec:Widths1}. The full
width of the structure $\mathcal{M}_{\mathrm{c}}$ saturated by the
aforementioned modes is determined in section \ref{sec:Widths2}. We make our
conclusions in the last part of the article \ref{sec:Conc}.

%%%%%%%%%%%%%%%%%%%%%%%%%%%%%%%%%%%%%%%%%%%%%%%%%%%%%%%%%%%%%%%%%

\section{Masses and current couplings of the molecules $\mathcal{M}_{\mathrm{%
b}}$ and $\mathcal{M}_{\mathrm{c}}$}

\label{sec:Mass}
%%%%%%%%%%%%%%%%%%%%%%%%%%%%%%%%%%%%%%%%%%%%%%%%%%%%%%%%%%%

Here, we consider the masses and current couplings of the molecules $%
\mathcal{M}_{\mathrm{b}}$ and $\mathcal{M}_{\mathrm{c}}$ in the framework of
the two-point QCD sum rule method. To this end, we employ the interpolating
currents for the molecules $\mathcal{M}_{\mathrm{b}}$ and $\mathcal{M}_{%
\mathrm{c}}$ and compute corresponding correlation functions.

Here we give, in details, calculations of $\mathcal{M}_{\mathrm{b}}$
molecule's spectroscopic parameters, but provide only results obtained for
the structure $\mathcal{M}_{\mathrm{c}}$. The molecule $\mathcal{M}_{\mathrm{%
b}}=\eta _{b}B_{c}^{-}$ with quark content $bb\overline{b}\overline{c}$ is
interpolated by the current $J(x)$,

\begin{equation}
J(x)=\overline{b}_{a}(x)i\gamma _{5}b_{a}(x)\overline{c}_{b}(x)i\gamma
_{5}b_{b}(x),  \label{eq:CR1}
\end{equation}%
where $a$ and $b$ are the color indices.

The scalar molecule $\mathcal{M}_{\mathrm{c}}=\eta _{c}B_{c}^{+}$ has the
similar current
\begin{equation}
\widetilde{J}(x)=\overline{c}_{a}(x)i\gamma _{5}c_{a}(x)\overline{b}%
_{b}(x)i\gamma _{5}c_{b}(x).  \label{eq:CR2}
\end{equation}

%%%%%%%%%%%%%%%%%%%%%%%%%%%%%%%%%%%%%%%%%%%%%%%%%%%%%%%%%%%%%%%%%%%%%%%%%%%%

\subsection{Parameters of the molecule $\mathcal{M}_{\mathrm{b}}$}

%%%%%%%%%%%%%%%%%%%%%%%%%%%%%%%%%%%%%%%%%%%%%%%%%%%%%%%%%%%%%%%%%%%%

To derive the SRs for the mass $m$ and current coupling $\Lambda $ of $%
\mathcal{M}_{\mathrm{b}}$, we explore the two-point correlation function
\begin{equation}
\Pi (p)=i\int d^{4}xe^{ipx}\langle 0|\mathcal{T}\{J(x)J^{\dag
}(0)\}|0\rangle ,  \label{eq:CF1}
\end{equation}%
where $\mathcal{T}$ is the time-ordered product of two currents.

In the sum rule approach this correlator has to be presented in two forms.
First, it should be expressed using the physical parameters $m$ and $\Lambda
$ of the molecule $\mathcal{M}_{\mathrm{b}}$. The correlator $\Pi ^{\mathrm{%
Phys}}(p)$ obtained by this way is, shortly, the physical side of the
required SRs. To find it, we take into account that $\Pi ^{\mathrm{Phys}}(p)$
is given by the formula
\begin{equation}
\Pi ^{\mathrm{Phys}}(p)=\frac{\langle 0|J|\mathcal{M}_{\mathrm{b}}\rangle
\langle \mathcal{M}_{\mathrm{b}}|J^{\dagger }|0\rangle }{m^{2}-p^{2}}+\cdots
,  \label{eq:Phys1}
\end{equation}%
and contains the contribution of the ground-state particle, as well as those
of the higher resonances and continuum states: The latter are shown in Eq.\ (%
\ref{eq:Phys1}) by the dots.

We tacitly assume in Eq.\ (\ref{eq:Phys1}) that the physical side of SR can
be approximated by a single pole term. But in the multiquark systems
two-meson reducible terms also contribute to $\Pi ^{\mathrm{Phys}}(p)$ \cite%
{Kondo:2004cr,Lee:2004xk}. The reason is that the current $J(x)$ interacts
not only with a molecule $\mathcal{M}_{\mathrm{b}}$, but also with the
relevant two-meson continuum. Such interaction, properly included into
analysis, leads to a finite width of the hadronic molecule and modifies the
quark propagator \cite{Agaev:2022ast}. It was numerously demonstrated that
two-meson contributions can be taken into account by rescaling the current
coupling $\Lambda $ but keeping stable the mass $m$ of an exotic four-quark
meson of interest. Computations prove that these effects are small and do
not overshoot ambiguities of SR analysis: Even for diquark-antidiquark
systems with the widths of a few hundred $\mathrm{MeV}$ corresponding
modifications amount to additional $5-7\%$ ambiguities in the current
couplings \cite{Agaev:2018vag,Sundu:2018nxt}, whereas SRs generate
uncertainties around of $10\%$ and higher. For the molecules $\mathcal{M}_{%
\mathrm{c}}$ and $\mathcal{M}_{\mathrm{b}}$ with the widths $70-90~\mathrm{%
MeV}$ the two-meson contaminations most likely are small, therefore in $\Pi
^{\mathrm{Phys}}(p)$ we use the zero-width single-pole approximation.

We rewrite $\Pi ^{\mathrm{Phys}}(p)$ using the matrix element
\begin{equation}
\langle 0|J|\mathcal{M}_{\mathrm{b}}\rangle =\Lambda ,  \label{eq:ME1}
\end{equation}%
and get
\begin{equation}
\Pi ^{\mathrm{Phys}}(p)=\frac{\Lambda ^{2}}{m^{2}-p^{2}}+\cdots .
\label{eq:Phys2}
\end{equation}%
The term $\Lambda ^{2}/(m^{2}-p^{2})$ is the invariant amplitude $\Pi ^{%
\mathrm{Phys}}(p^{2})$ required for following analysis.

Second, $\Pi (p)$ is calculated in the operator product expansion ($\mathrm{%
OPE}$) by employing heavy quark propagators. The result of these computations%
\begin{eqnarray}
&&\Pi ^{\mathrm{OPE}}(p)=i\int d^{4}xe^{ipx}\mathrm{Tr}\left\{ \left[ \gamma
_{5}S_{b}^{aa^{\prime }}(x)\gamma _{5}S_{b}^{a^{\prime }a}(-x)\right] \right.
\notag \\
&&\times \mathrm{Tr}\left[ \gamma _{5}S_{b}^{bb^{\prime }}(x)\gamma
_{5}S_{c}^{b^{\prime }b}(-x)\right] -\mathrm{Tr}\left[ \gamma
_{5}S_{b}^{ab^{\prime }}(x)\gamma _{5}S_{c}^{b^{\prime }b}(-x)\right.  \notag
\\
&&\left. \left. \times \gamma _{5}S_{b}^{ba^{\prime }}(x)\gamma
_{5}S_{b}^{a^{\prime }a}(-x)\right] \right\} ,  \label{eq:QCD1}
\end{eqnarray}%
is the QCD side $\Pi ^{\mathrm{OPE}}(p)$ of the sum rules, where $%
S_{b(c)}^{ab}(x)$ are the propagators of $b$ and $c$ quarks \cite%
{Agaev:2020zad}.

The function $\Pi ^{\mathrm{OPE}}(p)$ has also the simple Lorentz structure:
We label as $\Pi ^{\mathrm{OPE}}(p^{2})$ the corresponding invariant
amplitude. By equating two formulas for the amplitudes and applying the
assumption about the hadron-quark duality, and performing some
manipulations, we get the SRs for $m$ and $\Lambda $ (for further details
see, for example, Ref.\ \cite{Agaev:2024uza})
\begin{equation}
m^{2}=\frac{\Pi ^{\prime }(M^{2},s_{0})}{\Pi (M^{2},s_{0})},  \label{eq:Mass}
\end{equation}%
and
\begin{equation}
\Lambda ^{2}=e^{m^{2}/M^{2}}\Pi (M^{2},s_{0}).  \label{eq:Coupl}
\end{equation}

In Eq.\ (\ref{eq:Mass}), we employ $\Pi ^{\prime }(M^{2},s_{0})=d\Pi
(M^{2},s_{0})/d(-1/M^{2})$. Here, $\Pi (M^{2},s_{0})$ is the amplitude $\Pi
^{\mathrm{OPE}}(p^{2})$ after the Borel transformation and continuum
subtraction procedures. The Borel transformation is necessary to suppress
contribution of higher resonances and continuum states. The continuum
subtraction allows us to remove the suppressed terms from the QCD side of
the relevant equality. As a result, $\Pi (M^{2},s_{0})$ acquires a
dependence on the Borel $M^{2}$ and continuum subtraction $s_{0}$
parameters, and has the form
\begin{equation}
\Pi (M^{2},s_{0})=\int_{(3m_{b}+m_{c})^{2}}^{s_{0}}ds\rho ^{\mathrm{OPE}%
}(s)e^{-s/M^{2}}+\Pi (M^{2}).  \label{eq:CorrF}
\end{equation}%
The spectral density $\rho ^{\mathrm{OPE}}(s)$ is found as an imaginary part
of the function $\Pi ^{\mathrm{OPE}}(p^{2})$. In the current paper we
consider perturbative and dimension-four $\sim \langle \alpha _{s}G^{2}/\pi
\rangle $ and six $\sim \langle g_{s}^{3}G^{3}\rangle $ contributions to $%
\Pi ^{\mathrm{OPE}}(p^{2})$, therefore $\rho ^{\mathrm{OPE}}(s)$ contains
terms $\rho ^{\mathrm{pert.}}(s)$, $\rho ^{\mathrm{Dim4}}(s)$, and $\rho ^{%
\mathrm{Dim6}}(s)$. The nonperturbative function $\Pi (M^{2})$ is calculated
directly from the correlator $\Pi ^{\mathrm{OPE}}(p)$ and embrace effects of
terms which are not included into the spectral density.

To carry out the numerical calculations, we have to fix the parameters in
Eqs.\ (\ref{eq:Mass}) and (\ref{eq:Coupl}). The $b$ and $c$ quarks' masses
and gluon condensates $\langle \alpha _{s}G^{2}/\pi \rangle $ and $\langle
g_{s}^{3}G^{3}\rangle $ are universal quantities. In the current article, we
employ
\begin{eqnarray}
&&m_{c}=(1.2730\pm 0.0046)~\mathrm{GeV},  \notag \\
&&m_{b}=(4.183\pm 0.007)~\mathrm{GeV},  \notag \\
&&\langle \alpha _{s}G^{2}/\pi \rangle =(0.012\pm 0.004)~\mathrm{GeV}^{4}.
\label{eq:GluonCond} \\
&&\langle g_{s}^{3}G^{3}\rangle =(0.57\pm 0.29)~\mathrm{GeV}^{6}
\end{eqnarray}%
Quark masses $m_{c}$ and $m_{b}$ are calculated in the $\overline{\mathrm{MS}%
}$ scheme \cite{PDG:2024}. The condensates $\langle \alpha _{s}G^{2}/\pi
\rangle $ and $\langle g_{s}^{3}G^{3}\rangle $ were estimated in Refs.\ \cite%
{Shifman:1978bx,Shifman:1978by,Narison:2015nxh} from studies of different
processes.

\begin{figure}[h]
\includegraphics[width=8.5cm]{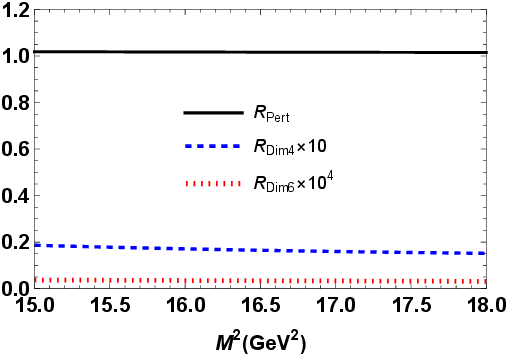}
\caption{Contributions of different terms to $\Pi (M^{2},s_{0})$ as
functions of $M^{2}$ at fixed $s_{0}=277.5~\mathrm{GeV}^{2}$.}
\label{fig:PCom}
\end{figure}

The parameters $M^{2}$ and $s_{0}$ depend on a analyzing problem and have to
satisfy standard restrictions of SR analyses. In the SR method the pole
contribution ($\mathrm{PC}$) should dominate in obtained quantities,
therefore, in computations we require fulfilment $\mathrm{PC}\geq 0.5$ .
Convergence of $\mathrm{OPE}$ is another condition for reliable SR studies.
In our case, the correlation function contains dimension-$4$ and -$6$ terms $%
\Pi ^{\mathrm{Dim4}}(M^{2},s_{0})$ and $\Pi ^{\mathrm{Dim6}}(M^{2},s_{0})$.
Then, the constraint $|\Pi ^{\mathrm{Dim4}}(M^{2},s_{0})+\Pi ^{\mathrm{Dim6}%
}(M^{2},s_{0})|\leq 0.05|\Pi (M^{2},s_{0})|$ is enough to ensure convergence
of $\mathrm{OPE}$. Last but not least is stability of final results upon
variations of $M^{2}$ and $s_{0}$.

\begin{figure}[h]
\includegraphics[width=8.5cm]{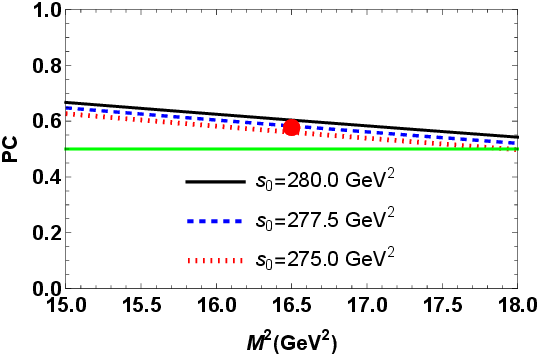}
\caption{Pole contribution $\mathrm{PC}$ as a function of $M^{2}$ at some $%
s_{0}$. The circle labels the point $M^{2}=16.5~\mathrm{GeV}^{2}$ and $%
s_{0}=277.5~\mathrm{GeV}^{2}$. }
\label{fig:PC}
\end{figure}

Numerical analysis is carried out over a broad range of the parameters $%
M^{2} $ and $s_{0}$. Collected results permits us to limit the working
regions for $M^{2}$ and $s_{0}$, where all standard conditions are
satisfied. We conclude that the intervals
\begin{equation}
M^{2}\in \lbrack 15,18]~\mathrm{GeV}^{2},\ s_{0}\in \lbrack 275,280]~\mathrm{%
GeV}^{2},  \label{eq:Wind1}
\end{equation}%
meet all these conditions. In fact, at maximal and minimal $M^{2}$ the pole
contribution averaged over $s_{0}$ is $\mathrm{PC}\approx 0.52$ and $\mathrm{%
PC}\approx 0.65$. The nonperturbative contributions are negative and at $%
M^{2}=15~\mathrm{GeV}^{2}$ $|\Pi ^{\mathrm{Dim4}}(M^{2},s_{0})+\Pi ^{\mathrm{%
Dim6}}(M^{2},s_{0})|$ constitutes approximately $2\%$ of the full result. In
Fig.\ \ref{fig:PCom} we plot the ratio
\begin{equation}
R_{\mathrm{pert.(N)}}(M^{2},s_{0})=\left\vert \frac{\Pi ^{\mathrm{pert.(DimN)%
}}(M^{2},s_{0})}{\Pi (M^{2},s_{0})}\right\vert ,
\end{equation}%
for the perturbative and nonperturbative components of $\Pi (M^{2},s_{0})$.
As is seen, contributions of the dimension-$6$ terms are negligibly small,
therefore in what follows we truncate $\mathrm{OPE}$ at dimension-$4$ terms.
The $\mathrm{PC}$ as a function of the Borel parameter is presented in Fig.\ %
\ref{fig:PC}, where all lines overshot the border $\mathrm{PC}=0.5$.

We calculate $m$ and $\Lambda $ as their mean values in the windows Eq.\ (%
\ref{eq:Wind1}) and get
\begin{eqnarray}
m &=&(15728\pm 90)~\mathrm{MeV},  \notag \\
\Lambda &=&(3.09\pm 0.32)~\mathrm{GeV}^{5}.  \label{eq:Result1}
\end{eqnarray}%
The predictions in Eq.\ (\ref{eq:Result1}) amount to SR results at $%
M^{2}=16.5~\mathrm{GeV}^{2}$ and $s_{0}=277.5~\mathrm{GeV}^{2}$, where $%
\mathrm{PC}\approx 0.58$, which guaranties the prevalence of $\mathrm{PC}$
in extracted quantities. The ambiguities in Eq.\ (\ref{eq:Result1}) are
formed due to choices of $M^{2}$ and $s_{0}$: Uncertainties connected with
errors in quark masses and gluon condensate are negligible.

The errors in Eq.\ (\ref{eq:Result1}) amount to $\pm 0.6\%$ of the mass $m$,
which proves the stability of this result. Uncertainties of $\Lambda $ are
larger and equal to $\pm 10\%$ remaining nevertheless inside borders
reasonable for the SR analysis. In Fig.\ \ref{fig:Mass}, we plot dependence
of $m$ on the parameters $M^{2}$ and $s_{0}$. For clear visualization of the
Borel platform the mass in the left panel of this figure is shown within
limits $M^{2}\in [12,20]~\mathrm{GeV}^{2}$.

\begin{widetext}

\begin{figure}[h!]
\begin{center}
\includegraphics[totalheight=6cm,width=8cm]{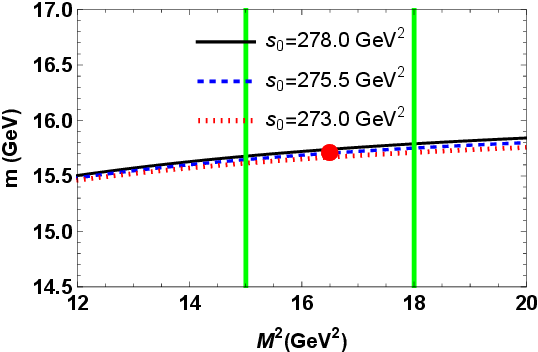}
\includegraphics[totalheight=6cm,width=8cm]{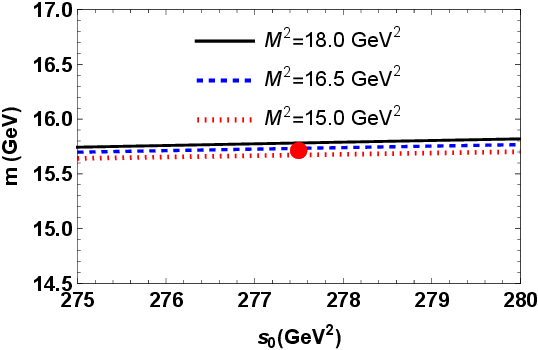}
\end{center}
\caption{Dependence of the mass $m$ on the parameters $M^{2}$ (left panel), and $s_0$ (right panel). Two vertical lines in the left panel limit a region, where $m$ has been extracted.}
\label{fig:Mass}
\end{figure}

\begin{figure}[h!]
\begin{center}
\includegraphics[totalheight=6cm,width=8cm]{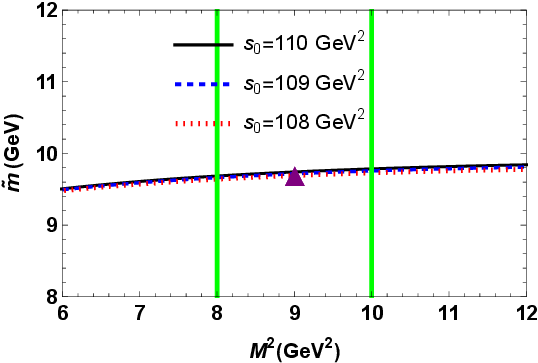}
\includegraphics[totalheight=6cm,width=8cm]{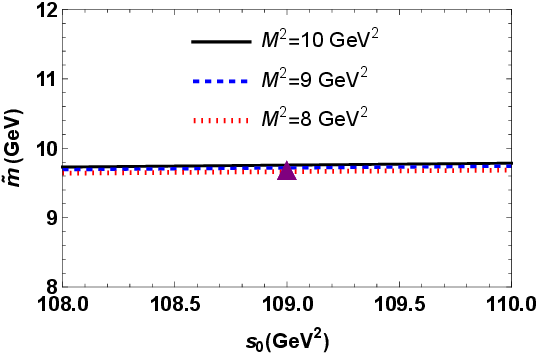}
\end{center}
\caption{Mass $\widetilde{m}$ as a function on the parameters $M^{2}$ (left panel), and $s_0$ (right panel). The triangle shows point $M^{2}=9~\mathrm{GeV}^{2}$ and
$s_{0}=109~\mathrm{GeV}^{2}$.}
\label{fig:Mass1}
\end{figure}

\end{widetext}

%%%%%%%%%%%%%%%%%%%%%%%%%%%%%%%%%%%%%%%%%%%%%%%%%%%%%%%%%%%%%%%%%

\subsection{Mass and current coupling of the molecule $\mathcal{M}_{\mathrm{c%
}}$}

%%%%%%%%%%%%%%%%%%%%%%%%%%%%%%%%%%%%%%%%%%%%%%%%%%%%%%%%%%%

The correlators $\widetilde{\Pi }^{\mathrm{Phys}}(p)$ and $\widetilde{\Pi }^{%
\mathrm{OPE}}(p)$, and SRs for parameters $\widetilde{m}$ and $\widetilde{%
\Lambda }$ of the molecule $\mathcal{M}_{\mathrm{c}}=\eta _{c}B_{c}^{+}$ do
not differ considerably from those of $\mathcal{M}_{\mathrm{b}}$. Therefore,
it is enough to present windows for $M^{2}$ and $s_{0}$. Numerical
calculations demonstrate that
\begin{equation}
M^{2}\in \lbrack 8,10]~\mathrm{GeV}^{2},\ s_{0}\in \lbrack 108,110]~\mathrm{%
GeV}^{2},  \label{eq:Wind1A}
\end{equation}%
satisfy all restrictions. Indeed, at maximal $M^{2}=10~\mathrm{GeV}^{2}$ the
pole contribution is $\mathrm{PC}\approx 0.50$, while at $M^{2}=8~\mathrm{GeV%
}^{2}$ it amounts o $\mathrm{PC}\approx 0.75$. The nonperturbative
contribution at $M^{2}=8~\mathrm{GeV}^{2}$ constitutes $2\%$ of the full
result.

The mass $\widetilde{m}$ and current coupling $\widetilde{\Lambda }$ of the
molecule $\mathcal{M}_{\mathrm{c}}$ are
\begin{eqnarray}
\widetilde{m} &=&(9712\pm 72)~\mathrm{MeV},  \notag \\
\widetilde{\Lambda } &=&(5.11\pm 0.48)\times 10^{-1}~\mathrm{GeV}^{5}.
\label{eq:Result2}
\end{eqnarray}%
These predictions effectively amount to the sum rule results at $M^{2}=9~%
\mathrm{GeV}^{2}$ and $s_{0}=109~\mathrm{GeV}^{2}$, where $\mathrm{PC}%
\approx 0.62$. The mass $\widetilde{m}$ as a function of the Borel and
continuum subtraction parameters $M^{2}$ and $s_{0}$ is depicted in Fig.\ %
\ref{fig:Mass1}.

%%%%%%%%%%%%%%%%%%%%%%%%%%%%%%%%%%%%%%%%%%%%%%%%%%%%%%%%%%%%%%%%%

\section{Full decay width of $\mathcal{M}_{\mathrm{b}}$}

\label{sec:Widths1}

%%%%%%%%%%%%%%%%%%%%%%%%%%%%%%%%%%%%%%%%%%%%%%%%%%%%%%%%%%%
In this section we calculate the full decay width of the hadronic molecule $%
\mathcal{M}_{\mathrm{b}}$. Information on the mass of $\mathcal{M}_{\mathrm{b%
}}$ permits one to find its decay channels. The process $\mathcal{M}_{%
\mathrm{b}}\rightarrow \eta _{b}B_{c}^{-}$ is kinematically allowed decay
channel of $\mathcal{M}_{\mathrm{b}}$. In fact, the masses $m_{\eta
_{b}}=(9398.7\pm 2.0)~\mathrm{MeV}$ and $m_{B_{c}}=(6274.47\pm 0.27\pm 0.17)~%
\mathrm{MeV}$ \cite{PDG:2024} of the final-state mesons establish the
threshold $15673~\mathrm{MeV}$ which is less than central value of $m=15728~%
\mathrm{MeV}$. This is the dominant mode of $\mathcal{M}_{\mathrm{b}}$,
because all of its valence quarks appear in the final-state particles.

It is interesting that $m=15728~\mathrm{MeV}$ is rather close to $\eta
_{b}B_{c}^{-}$ threshold and, due to uncertainties in estimation of $m$, in
its lower value $m=15638~\mathrm{MeV}$ lies below it. In other words the
scalar molecule $\mathcal{M}_{\mathrm{b}}$ can be considered as a bound
state of the mesons $\eta _{b}$ and $B_{c}^{-}$. Then the dominant decay
channel $\mathcal{M}_{\mathrm{b}}\rightarrow \eta _{b}B_{c}^{-}$ becomes
kinematically forbidden for $\mathcal{M}_{\mathrm{b}}$. But this does not
mean that it is stable against strong-interaction decays. In fact, there is
an alternative mechanism for its transformation to ordinary particles.

The alternative decay channels of the molecule $\mathcal{M}_{\mathrm{b}}$
are ones generated by annihilation of $b\overline{b}$ quarks in $\mathcal{M}%
_{\mathrm{b}}$ to $q\overline{q}$ and $s\overline{s}$ pairs. Afterwards,
initial $b$ and $\overline{c}$ quarks from $\mathcal{M}_{\mathrm{b}}$ and
light quarks form pairs of $B_{(s)}^{(\ast )}D_{(s)}^{(\ast )}$ mesons with
appropriate quantum numbers and charges. In the SR method in corresponding
correlation functions we relate the vacuum expectation value $\langle
\overline{b}b\rangle $\ of $b$ quarks to the gluon condensate $\langle
\alpha _{s}G^{2}/\pi \rangle $, therefore these processes are subleading
modes of the molecule $\mathcal{M}_{\mathrm{b}}$. Nevertheless, total
contribution of such channel to the full decay width of $\mathcal{M}_{%
\mathrm{b}}$ may be sizeable. Here, we are going to take into account decays
to mesons $B^{-}\overline{D}^{0}$, $\overline{B}^{0}D^{-}$, $\overline{B}%
_{s}^{0}D_{s}^{-}$, $B^{\ast -}\overline{D}^{\ast 0}$, $\overline{B}^{\ast
0}D^{\ast -}$, and $\overline{B}_{s}^{\ast 0}D_{s}^{\ast -}$.

%%%%%%%%%%%%%%%%%%%%%%%%%%%%%%%%%%%%%%%%%%%%%%%%%%%%%%%%%%%%%%%%%

\subsection{Process $\mathcal{M}_{\mathrm{b}}\rightarrow \protect\eta %
_{b}B_{c}^{-}$}

%%%%%%%%%%%%%%%%%%%%%%%%%%%%%%%%%%%%%%%%%%%%%%%%%%%%%%%%%%%

The width of the process $\mathcal{M}_{\mathrm{b}}(p)\rightarrow \eta
_{b}(p^{\prime })B_{c}^{-}(q)$ besides the known parameters depends on the
strong coupling $g$ at the vertex $\mathcal{M}_{\mathrm{b}}\eta
_{b}B_{c}^{-} $. In its turn, $g$ can be computed at the mass shell $%
q^{2}=m_{B_{c}}^{2}$ using the form factor $g(q^{2})$. To evaluate $g(q^{2})$
we analyze the following three-point correlation function
\begin{eqnarray}
\Pi (p,p^{\prime }) &=&i^{2}\int d^{4}xd^{4}ye^{ip^{\prime
}y}e^{-ipx}\langle 0|\mathcal{T}\{J^{\eta _{b}}(y)  \notag \\
&&\times J^{B_{c}^{-}}(0)J^{\dagger }(x)\}|0\rangle ,  \label{eq:CF1a}
\end{eqnarray}%
with $J^{\eta _{b}}(x)$ and $J^{B_{c}^{-}}(x)$ being the currents which
interpolate the pseudoscalar mesons $\eta _{b}$ and $B_{c}^{-}$, and have
the forms
\begin{equation}
J^{\eta _{b}}(x)=\overline{b}_{i}(x)i\gamma _{5}b_{i}(x),\ J^{B_{c}^{-}}(x)=%
\overline{c}_{j}(x)i\gamma _{5}b_{j}(x).
\end{equation}%
Here, $i$ and $j$ are the color indices. The four-momentum $p$ of the
molecule $\mathcal{M}_{\mathrm{b}}$ is connected by the equality $%
p=p^{\prime }+q$ to momenta of mesons.

It is known that the correlator Eq.\ (\ref{eq:CF1a}) expressed using
parameters of particles $\mathcal{M}_{\mathrm{b}}$, $\eta _{b}$ and $%
B_{c}^{-}$ is the phenomenological side of SR $\Pi ^{\mathrm{Phys}%
}(p,p^{\prime })$. To find $\Pi ^{\mathrm{Phys}}(p,p^{\prime })$, we insert
into Eq.\ (\ref{eq:CF1a}) full system of intermediate states for the
particles $\mathcal{M}_{\mathrm{b}}$, $\eta _{b}$ and $B_{c}^{-}$ and carry
out four-integrals over $x$ and $y$. Having dissected the contribution of
the ground-state particles and using a naive factorization approximation, we
obtain
\begin{eqnarray}
&&\Pi ^{\mathrm{Phys}}(p,p^{\prime })=\frac{\langle 0|J^{\eta _{b}}|\eta
_{b}(p^{\prime })\rangle }{p^{\prime 2}-m_{\eta _{b}}^{2}}\frac{\langle
0|J^{B_{c}^{-}}|B_{c}^{-}(q)\rangle }{q^{2}-m_{B_{c}}^{2}}  \notag \\
&&\times \langle \eta _{b}(p^{\prime })B_{c}^{-}(q)|\mathcal{M}_{\mathrm{b}%
}(p)\rangle \frac{\langle \mathcal{M}_{\mathrm{b}}(p)|J^{\dagger }|0\rangle
}{p^{2}-m^{2}}  \notag \\
&&+\cdots .  \label{eq:TP1}
\end{eqnarray}%
The ellipses above denote effects of excited and continuum states.

By applying to Eq.\ (\ref{eq:TP1}) the matrix elements of the mesons $\eta
_{b}$ and $B_{c}^{-}$%
\begin{eqnarray}
\langle 0|J^{\eta _{b}}|\eta _{b}(p^{\prime })\rangle &=&\frac{f_{\eta
_{b}}m_{\eta _{b}}^{2}}{2m_{b}},  \notag \\
\langle 0|J^{B_{c}^{-}}|B_{c}^{-}(q)\rangle &=&\frac{f_{B_{c}}m_{B_{c}}^{2}}{%
m_{b}+m_{c}},  \label{eq:ME1A}
\end{eqnarray}%
one can simplify $\Pi ^{\mathrm{Phys}}$. Above, $f_{\eta _{b}}$and $%
f_{B_{c}} $ are the decay constants of the corresponding mesons. We have to
introduce also a formula for the vertex $\langle \eta _{b}(p^{\prime
})B_{c}^{-}(q)|\mathcal{M}_{\mathrm{b}}(p)\rangle $. It has a simple form
\begin{equation}
\langle \eta _{b}(p^{\prime })B_{c}^{-}(q)|\mathcal{M}_{\mathrm{b}%
}(p)\rangle =g(q^{2})p\cdot p^{\prime }.  \label{eq:SAVAV}
\end{equation}%
As a result, we get
\begin{eqnarray}
&&\Pi ^{\mathrm{Phys}}(p,p^{\prime })=g(q^{2})\frac{\Lambda f_{\eta
_{b}}m_{\eta _{b}}^{2}f_{B_{c}}m_{B_{c}}^{2}}{2m_{b}(m_{b}+m_{c})\left(
p^{2}-m^{2}\right) }  \notag \\
&&\times \frac{1}{(p^{\prime 2}-m_{\eta _{b}}^{2})(q^{2}-m_{B_{c}}^{2})}%
\frac{m^{2}+m_{\eta _{b}}^{2}-q^{2}}{2}+\cdots .  \notag \\
&&  \label{eq:PhysS1}
\end{eqnarray}%
This is the invariant amplitude $\Pi ^{\mathrm{Phys}}(p^{2},p^{\prime
2},q^{2})$ which will be used to obtain SR for $g(q^{2})$.

The correlator $\Pi (p,p^{\prime })$ computed in terms of quark propagators
reads
\begin{eqnarray}
&&\Pi ^{\mathrm{OPE}}(p,p^{\prime })=\int d^{4}xd^{4}ye^{ip^{\prime
}y}e^{-ipx}\left\{ \mathrm{Tr}\left[ \gamma _{5}S_{b}^{ia}(y-x)\right.
\right.  \notag \\
&&\left. \times \gamma _{5}S_{b}^{ai}(x-y)\right] \mathrm{Tr}\left[ \gamma
_{5}S_{b}^{jb}(-x)\gamma _{5}S_{c}^{bj}(x)\right]  \notag \\
&&\left. -\mathrm{Tr}\left[ \gamma _{5}S_{b}^{ib}(y-x)\gamma
_{5}S_{c}^{bj}(x)\gamma _{5}S_{b}^{ja}(-x)\gamma _{5}S_{b}^{ai}(x-y)\right]
\right\} .  \notag \\
&&  \label{eq:CF3}
\end{eqnarray}%
The correlator $\Pi ^{\mathrm{OPE}}(p,p^{\prime })$ has a simple Lorentz $%
\sim \mathrm{I}$ organization as well, and is equal to the amplitude $\Pi ^{%
\mathrm{OPE}}(p^{2},p^{\prime 2},q^{2})$. In the present work, this
amplitude is calculated by taking into account $\mathrm{Dim4}$ terms $\sim
\langle \alpha _{s}G^{2}/\pi \rangle $.

Having equated $\Pi ^{\mathrm{Phys}}(p^{2},p^{\prime 2},q^{2})$ and $\Pi ^{%
\mathrm{OPE}}(p^{2},p^{\prime 2},q^{2})$, performed the double Borel
transformations over the variables $-p^{2}$, $-p^{\prime 2}$ and under the
quark-hadron duality assumption subtracted contributions of excited and
continuum states from the QCD side of this equality, we derive the sum rule
for $g(q^{2})$
\begin{eqnarray}
&&g(q^{2})=\frac{4m_{b}(m_{b}+m_{c})(q^{2}-m_{B_{c}}^{2})}{\Lambda f_{\eta
_{b}}m_{\eta _{b}}^{2}f_{B_{c}}m_{B_{c}}^{2}(m^{2}+m_{\eta _{b}}^{2}-q^{2})}
\notag \\
&&\times e^{m^{2}/M_{1}^{2}}e^{m_{\eta _{b}}^{2}/M_{2}^{2}}\Pi (\mathbf{M}%
^{2},\mathbf{s}_{0},q^{2}).  \label{eq:SRG}
\end{eqnarray}%
Here $\Pi (\mathbf{M}^{2},\mathbf{s}_{0},q^{2})$ is given by the expression
\begin{eqnarray}
\Pi (\mathbf{M}^{2},\mathbf{s}_{0},q^{2})
&=&\int_{(3m_{b}+m_{c})^{2}}^{s_{0}}\int_{4m_{b}^{2}}^{s_{0}^{\prime
}}dsds^{\prime }e^{-s/M_{1}^{2}}  \notag \\
&&\times e^{-s^{\prime }/M_{2}^{2}}\rho (s,s^{\prime },q^{2}),
\end{eqnarray}%
where the spectral density $\rho (s,s^{\prime },q^{2})$ amounts to the
imaginary part of $\Pi ^{\mathrm{OPE}}(s,s^{\prime },q^{2})$.

The correlator $\Pi (\mathbf{M}^{2},\mathbf{s}_{0},q^{2})$ depends on the
parameters $\mathbf{M}^{2}=(M_{1}^{2},M_{2}^{2})$ and $\mathbf{s}%
_{0}=(s_{0},s_{0}^{\prime })$ where the pairs $(M_{1}^{2},s_{0})$ and $%
(M_{2}^{2},s_{0}^{\prime })$ are related to $\mathcal{M}_{\mathrm{b}}$ and $%
\eta _{b}$ channels. Restrictions imposed on $\mathbf{M}^{2}$ and $\mathbf{s}%
_{0}$ are standard in SR calculations and have been detailed above (see,
Sec. \ref{sec:Mass}). Our analysis demonstrates that Eq.\ (\ref{eq:Wind1})
for the parameters $(M_{1}^{2},s_{0})$ and
\begin{equation}
M_{2}^{2}\in \lbrack 9,11]~\mathrm{GeV}^{2},\ s_{0}^{\prime }\in \lbrack
95,99]~\mathrm{GeV}^{2}.  \label{eq:Wind3}
\end{equation}%
for $(M_{2}^{2},s_{0}^{\prime })$ meet these requirements. The mass and
decay constant of the mesons $\eta _{b}$ and $B_{c}^{-}$ necessary for
numerical computations, as well as parameters of particles that emerge while
studying other decays are collected in Table\ \ref{tab:Param}. The
parameters of the $B_{c}^{\ast }$ meson are model-dependent predictions \cite%
{Godfrey:2004ya,Eichten:2019gig}. Other masses are borrowed from Ref.\ \cite%
{PDG:2024}, while decay constants were extracted from experimental
measurements or computed using various theoretical methods \cite%
{Davies:2021lkj,Wang:2024fwc,Veliev:2010vd,
Lakhina:2006vg,Rosner:2015wva,Lucha:2014spa,Lubicz:2016bbi,Narison:2012xy,Chang:2018aut}%
.

\begin{table}[tbp]
\begin{tabular}{|c|c|c|}
\hline\hline
Mesons & mass ($\mathrm{MeV}$) & D.C. ($\mathrm{MeV}$) \\ \hline\hline
$\eta_{b}$ & $9398.7 \pm 2.0$ & $724 \pm 12$\ \cite{Davies:2021lkj} \\
$B_{c}^{\pm}$ & $6274.47 \pm 0.27 \pm 0.17$ & $371 \pm 37$ \cite%
{Wang:2024fwc} \\
$B_{c}^{\ast \pm}$ & $6338$\ \cite{Godfrey:2004ya} & $471$\ \cite%
{Eichten:2019gig} \\
$\eta_{c}$ & $2984.1 \pm 0.4$ & $421 \pm 35$\ \cite{Veliev:2010vd} \\
$J/\psi$ & $3096.900 \pm 0.006$ & $411 \pm 7$\ \cite{Lakhina:2006vg} \\
$\overline{D}^0$ & $1864.84 \pm 0.05$ & $211.9 \pm 1.1$\ \cite%
{Rosner:2015wva} \\
$D^{\pm}$ & $1869.66 \pm 0.05$ & $211.9 \pm 1.1$\ \cite{Rosner:2015wva} \\
$D_{s}^{\pm}$ & $1968.35 \pm 0.07$ & $249.0 \pm 1.2$ \cite{Rosner:2015wva}
\\
$\overline{D}^{\ast 0}$ & $2006.85 \pm 0.05$ & $252.2 \pm 22.66$\ \cite%
{Lucha:2014spa} \\
$D^{\ast \pm}$ & $2010.26 \pm 0.05 $ & $252.2 \pm 22.66$\ \cite%
{Lucha:2014spa} \\
$D_{s}^{\ast \pm}$ & $2112.2 \pm 0.4$ & $268.8 \pm 6.6$\ \cite%
{Lubicz:2016bbi} \\
$\overline{B}^{0}$ & $5279.72 \pm 0.08$ & $206 \pm 7$\ \cite{Narison:2012xy}
\\
$B^{\pm}$ & $5279.41 \pm 0.07$ & $206 \pm 7$\ \cite{Narison:2012xy} \\
$\overline{B}_{s}^{0}$ & $5366.93 \pm 0.10 $ & $234 \pm 5$\ \cite%
{Narison:2012xy} \\
$\overline{B}^{\ast 0}$, $B^{\ast \pm}$ & $5324.75 \pm 0.20$ & $210 \pm 6$\
\cite{Chang:2018aut} \\
$\overline{B}_{s}^{\ast 0}$ & $5415.4 \pm 1.4$ & $221 \pm 7$\ \cite%
{Chang:2018aut} \\ \hline\hline
\end{tabular}%
\caption{Masses and decay constants (D.C.) of the mesons that appear in
decays of the hadronic molecules $\mathcal{M}_{b}$ and $\mathcal{M}_{c}$. }
\label{tab:Param}
\end{table}

\begin{figure}[h]
\includegraphics[width=8.5cm]{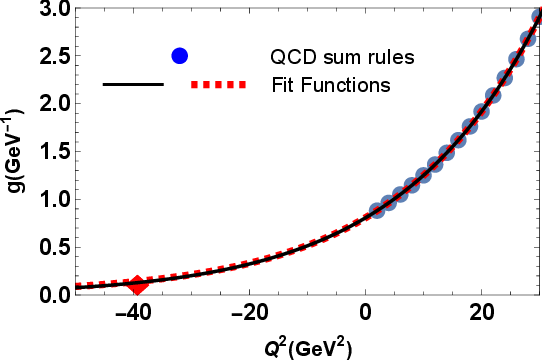}
\caption{The sum rule's data and extrapolating functions $\mathcal{G}%
(Q^{2},m^2)$ (solid line) and $\mathcal{G}_{A}(Q^{2},m^2)$. The diamond is
placed at $Q^{2}=-m_{B_c}^{2}$. }
\label{fig:Fit}
\end{figure}

The SR method leads to credible results in the Euclidean region $q^{2}<0$.
At the same time, $g(q^{2})$ becomes equal to $g$ at the mass shell $%
q^{2}=m_{B_{c}}^{2}$. For this reason, we use the function $g(Q^{2})$ with $%
Q^{2}=-q^{2}$ and utilize it in following analysis. The SR predictions for $%
g(Q^{2})$ are shown in Fig.\ \ref{fig:Fit}, where $Q^{2}$ changes within
borders $Q^{2}=2-30~\mathrm{GeV}^{2}$.

To extract $g$ at the mass shell $q^{2}=-Q^{2}=m_{B_{c}}^{2}$, we employ the
extrapolating function $\mathcal{G}(Q^{2},m^{2})$ which at $Q^{2}>0$
coincides with SR data, but can also be applied in the domain $Q^{2}<0$.
This function has the analytical form
\begin{equation}
\mathcal{G}_{i}(Q^{2},m^{2})=\mathcal{G}_{i}^{0}\mathrm{\exp }\left[
c_{i}^{1}\frac{Q^{2}}{m^{2}}+c_{i}^{2}\left( \frac{Q^{2}}{m^{2}}\right) ^{2}%
\right] ,  \label{eq:FitF}
\end{equation}%
where $\mathcal{G}_{i}^{0}$, $c_{i}^{1}$, and $c_{i}^{2}$ are constants
obtained from comparison with SR data. Then, it is not difficult to find
\begin{equation}
\mathcal{G}^{0}=0.81~\mathrm{GeV}^{-1},c^{1}=10.99,\text{and }c^{2}=-3.46.
\label{eq:FF1}
\end{equation}%
In Fig.\ \ref{fig:Fit} we plot $\mathcal{G}(Q^{2},m^{2})$ as well: Nice
agreement of $\mathcal{G}(Q^{2},m^{2})$ and SR data is evident. Then, for $g$
we obtain
\begin{equation}
g\equiv \mathcal{G}(-m_{B_{c}}^{2},m^{2})=(1.3\pm 0.3)\times 10^{-1}\
\mathrm{GeV}^{-1}.  \label{eq:g1}
\end{equation}

This prediction has been obtained by applying the function Eq.\ (\ref%
{eq:FitF}). But, in general, the SR data can be extrapolated to region of $%
Q^{2}<0$ by means of alternative fit functions. To study this effect, we use
the new function
\begin{equation}
\mathcal{G}_{A}(Q^{2},m^{2})=\frac{l_{0}\left( 1-Q^{2}/m^{2}\right) ^{-l_{1}}%
}{1-l_{2}(Q^{2}/m^{2})+l_{3}\left( Q^{4}/m^{4}\right) },  \label{eq:FitF2}
\end{equation}
where $l_{0}$, $l_{1}$, $l_{2}$ and $l_{3}$ are fitting parameters. Having
compared Eq. (\ref{eq:FitF2}) and SR data we find $l_{0}=0.807~\mathrm{GeV}%
^{-1}$, $l_{1}=9.013$, $l_{2}=1.910$ and $l_{3}=8.274$. In Fig.\ \ref%
{fig:Fit}\ we plot $\mathcal{G}_{A}(Q^{2},m^{2})$, in which one sees its
perfect agreement with SR data and $\mathcal{G}(Q^{2},m^{2})$. The function $%
\mathcal{G}_{A}(Q^{2},m^{2})$ predicts for the the strong coupling $g=0.14~%
\mathrm{GeV}^{-1}$. The deviation $|0.01|$ of this value from one presented
in Eq.\ (\ref{eq:g1}) is a factor $3$ smaller than uncertainties $\pm 0.03$
of $g$ generated by the sum rule method itself. Therefore, here and in what
follows, we use Eq.\ (\ref{eq:FitF}) and neglect ambiguities connected with
a choice of different extrapolating functions.

The width of the decay $\mathcal{M}_{\mathrm{b}}\rightarrow \eta
_{b}B_{c}^{-}$ is given by the formula%
\begin{equation}
\Gamma \left[ \mathcal{M}_{\mathrm{b}}\rightarrow \eta _{b}B_{c}^{-}\right]
=g^{2}\frac{m_{\eta _{b}}^{2}\lambda }{8\pi }\left( 1+\frac{\lambda ^{2}}{%
m_{\eta _{b}}^{2}}\right) ,  \label{eq:PDw2}
\end{equation}%
where $\lambda =\lambda (m,m_{\eta _{b}},m_{B_{c}})$ and
\begin{equation}
\lambda (a,b,c)=\frac{\sqrt{%
a^{4}+b^{4}+c^{4}-2(a^{2}b^{2}+a^{2}c^{2}+b^{2}c^{2})}}{2a}.
\end{equation}

Then, we obtain $\ $%
\begin{equation}
\Gamma \left[ \mathcal{M}_{\mathrm{b}}\rightarrow \eta _{b}B_{c}^{-}\right]
=(37.8\pm 15.4)~\mathrm{MeV}.  \label{eq:DW2}
\end{equation}%
The error above is generated by the ambiguities of the coupling $g$ and the
masses of the particles $\mathcal{M}_{\mathrm{b}}$ (upper limit), $\eta _{b}$
and $B_{c}^{-}$.

%%%%%%%%%%%%%%%%%%%%%%%%%%%%%%%%%%%%%%%%%%%%%%%%%%%%%%%%%%%%%%%%%

\subsection{Decays of $\mathcal{M}_{\mathrm{b}}$ triggered by $\overline{b}b$
annihilation}

%%%%%%%%%%%%%%%%%%%%%%%%%%%%%%%%%%%%%%%%%%%%%%%%%%%%%%%%%%%

As it has been explained, annihilation of $\overline{b}b$ quarks gives rise
to numerous decay channels of the molecule $\mathcal{M}_{\mathrm{b}}$. The
processes $\mathcal{M}_{\mathrm{b}}\rightarrow B^{-}\overline{D}^{0}$, $%
\overline{B}^{0}D^{-}$, $\overline{B}_{s}^{0}D_{s}^{-}$, and vector $B^{\ast
-}\overline{D}^{\ast 0}$, $\overline{B}^{\ast 0}D^{\ast -}$, $\overline{B}%
_{s}^{\ast 0}D_{s}^{\ast -}$are among these modes. Let us first consider the
decays to pairs of pseudoscalar mesons. In our present studies we adopt the
approximations $m_{\mathrm{u}}=m_{\mathrm{d}}=0$ and $m_{\mathrm{s}%
}=(93.5\pm 0.8)~\mathrm{MeV}$. The correlation functions for the decays $%
\mathcal{M}_{\mathrm{b}}\rightarrow B^{-}\overline{D}^{0}$ and $\mathcal{M}_{%
\mathrm{b}}\rightarrow \overline{B}^{0}D^{-}$ contain $u$ and $d$ quark
propagators which are the same in this approximation. The partial widths of
these processes may differ from each other due to parameters of the
particles involved into decays. We use the same decay constants for the
neutral and charged mesons, therefore their masses are only possible sources
of potential variations. From Table\ \ref{tab:Param} it is seen that
differences between the masses of the mesons $B^{-}$and $\overline{B}^{0}$,
as well as $\overline{D}^{0}$ and $D^{-}$ ones are very small. For this
reason, we calculate the partial width $\Gamma \left[ \mathcal{M}_{\mathrm{b}%
}\rightarrow B^{-}\overline{D}^{0}\right] $ of the decay $\mathcal{M}_{%
\mathrm{b}}\rightarrow B^{-}\overline{D}^{0}$, and employ an approximate
relation $\Gamma \left[ \mathcal{M}_{\mathrm{b}}\rightarrow \overline{B}%
^{0}D^{-}\right] \approx \Gamma \left[ \mathcal{M}_{\mathrm{b}}\rightarrow
B^{-}\overline{D}^{0}\right] $. The similar arguments are valid in the case
of the decays to vector mesons as well.

Let us consider the decay $\mathcal{M}_{\mathrm{b}}\rightarrow B^{-}%
\overline{D}^{0}$ of the molecule $\mathcal{M}_{\mathrm{b}}$ in a detailed
form. Our aim is to extract the strong coupling $g$ at the vertex $\mathcal{M%
}_{\mathrm{b}}B^{-}\overline{D}^{0}$. To this end, we investigate the
three-point correlator
\begin{eqnarray}
\Pi _{1}(p,p^{\prime }) &=&i^{2}\int d^{4}xd^{4}ye^{ip^{\prime
}y}e^{-ipx}\langle 0|\mathcal{T}\{J^{B^{-}}(y)  \notag \\
&&\times J^{\overline{D}^{0}}(0)J^{\dagger }(x)\}|0\rangle ,  \label{eq:CF3A}
\end{eqnarray}%
where $J^{B^{-}}(x)$ and $J^{\overline{D}^{0}}(x)$ are currents for the
mesons $B^{-}$ and $\overline{D}^{0}$. They have the following forms
\begin{equation}
J^{B^{-}}(x)=\overline{u}_{i}(x)i\gamma _{5}b_{i}(x),\ J^{\overline{D}%
^{0}}(x)=\overline{c}_{j}(x)i\gamma _{5}u_{j}(x).
\end{equation}%
The matrix elements of these mesons employed to calculate the physical side
of the sum rule for the relevant form factor $g_{1}(q^{2})$ are
\begin{eqnarray}
\langle 0|J^{B^{-}}|B^{-}(p^{\prime })\rangle &=&\frac{f_{B}m_{B}^{2}}{m_{b}}%
,  \notag \\
\langle 0|J^{\overline{D}^{0}}|\overline{D}^{0}(q)\rangle &=&\frac{f_{D}m_{%
\overline{D}^{0}}^{2}}{m_{c}}.
\end{eqnarray}%
In formulas above $m_{B},\ m_{\overline{D}^{0}}$ and $f_{B},$ $f_{D}$ are
the masses and decay constants of the these particles. The vertex $\langle
B^{-}(p^{\prime })\overline{D}^{0}(q)|\mathcal{M}_{\mathrm{b}}(p)\rangle $
and correlator $\Pi _{1}^{\mathrm{Phys}}(p,p^{\prime })$ are similar to
those obtained in the previous subsection.

The QCD side of SR for the form factor $g_{1}(q^{2})$ is given by the
expression
\begin{eqnarray}
&&\Pi _{1}^{\mathrm{OPE}}(p,p^{\prime })=\frac{1}{3}\int
d^{4}xd^{4}ye^{ip^{\prime }y}e^{-ipx}\langle \overline{b}b\rangle  \notag \\
&&\times \mathrm{Tr}\left[ \gamma _{5}S_{b}^{ia}(y-x)S_{c}^{aj}(x)\gamma
_{5}S_{u}^{ji}(-y)\right] .  \label{eq:CF3AA}
\end{eqnarray}%
The correlation function $\Pi _{1}^{\mathrm{OPE}}(p,p^{\prime })$ contains
three quark propagators and vacuum condensate $\langle \overline{b}b\rangle $%
, and differs from standard correlators, for example, from Eq.\ (\ref{eq:CF3}%
), which depends on four quark propagators. To calculate $\Pi _{1}^{\mathrm{%
OPE}}(p,p^{\prime })$, as usual, we contract heavy and light quark fields.
Because the pair of mesons $B^{-}\overline{D}^{0}$ contains only one $b$%
-quark field, remaining pair of $\overline{b}b$ quarks in the molecule $%
\mathcal{M}_{\mathrm{b}}$ form a local vacuum condensate. This is a reason
for appearance of the factor $\langle \overline{b}b\rangle $ in Eq.\ (\ref%
{eq:CF3AA}). Stated differently, $\langle \overline{b}b\rangle $ emerges as
a substituent of a quark propagator and has to be treated on equal footing
with it.

To continue calculations, we employ the heavy and light quark propagators
and use their standard formulas. We utilize also the relation between the
heavy quark and gluon condensates to express $\Pi _{1}^{\mathrm{OPE}%
}(p,p^{\prime })$ in terms of known parameters. The available expression for
$\langle \overline{b}b\rangle $ reads \cite%
{Shifman:1978bx,Generalis:1983hb,Bagan:1985zp}
\begin{eqnarray}
m_{b}\langle \overline{b}b\rangle &=&-\frac{1}{12}\langle \frac{\alpha
_{s}G^{2}}{\pi }\rangle +\frac{1}{m_{b}^{2}}\langle \frac{\alpha _{s}G^{3}}{%
\pi }\rangle  \notag \\
&&\times \left( -\frac{1}{48}+\frac{13}{720}\right) +\cdots .
\label{eq:Conden}
\end{eqnarray}%
In our analysis we use only the first term in Eq.\ (\ref{eq:Conden})
extracted in Ref.\ \cite{Shifman:1978bx}, because next ones are suppressed
by additional powers of $m_{b}^{-1}$ and can be neglected.

\begin{figure}[h]
\includegraphics[width=8.5cm]{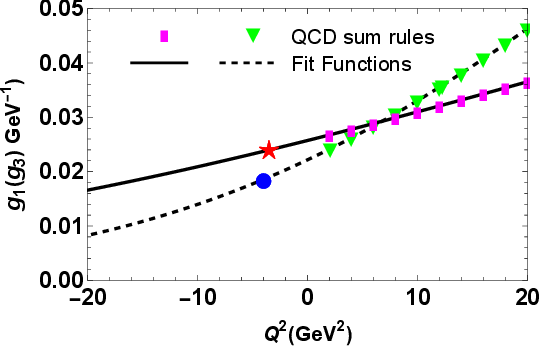}
\caption{The QCD data for the form factors $g_{1}(Q^{2})$ and $g_{3}(Q^{2})$
and fit functions $\mathcal{G}_{1}(Q^{2},m^{2})$ (solid line), $\mathcal{G}%
_{3}(Q^{2},m^{2})$ (dashed line). The red star and blue circle show
positions of the points $Q^{2}=-m_{\overline{D}^{0}}^{2}$ and $%
Q^{2}=-m_{D^{\ast }}^{2}$, respectively. }
\label{fig:Fit1}
\end{figure}

The form factor $g_{1}(Q^{2})$ is computed in the region $Q^{2}=2-20\
\mathrm{GeV}^{2}$. In numerical calculations for parameters $%
(M_{1}^{2},s_{0})$ we have used Eq.\ (\ref{eq:Wind1}), whereas $%
(M_{2}^{2},s_{0}^{\prime })$ have been chosen in the following intervals
\begin{equation}
M_{2}^{2}\in \lbrack 5.5,6.5]~\mathrm{GeV}^{2},\ s_{0}^{\prime }\in \lbrack
33.5,34.5]~\mathrm{GeV}^{2}.
\end{equation}%
Predictions obtained for $g_{1}(Q^{2})$ are displayed in Fig.\ \ref{fig:Fit1}%
. The extrapolating function $\mathcal{G}_{1}(Q^{2},m^{2})$ is fixed by the
constants $\mathcal{G}_{1}^{0}=0.026~\mathrm{GeV}^{-1}$, $c_{1}^{1}=4.88$,
and $c_{1}^{2}=-6.70.$ Then the coupling $g_{1}$ can be extracted at the
point $Q^{2}=-m_{\overline{D}^{0}}^{2}$ and is equal to%
\begin{equation}
g_{1}\equiv \mathcal{G}_{1}(-m_{\overline{D}^{0}}^{2},m^{2})=(2.42\pm
0.39)\times 10^{-2}\ \mathrm{GeV}^{-1}.
\end{equation}%
This leads to the following results for width of the decay $\mathcal{M}_{%
\mathrm{b}}\rightarrow B^{-}\overline{D}^{0}$%
\begin{equation}
\Gamma \left[ \mathcal{M}_{\mathrm{b}}\rightarrow B^{-}\overline{D}^{0}%
\right] =(11.9\pm 2.8)~\mathrm{MeV}.
\end{equation}%
Note that uncertainties in the width is total errors connected by
uncertainties both in $g_{1}$ and the masses $\mathcal{M}_{\mathrm{b}}$ , $%
m_{B}$ and $m_{\overline{D}^{0}}$.
The decay $\mathcal{M}_{\mathrm{b}}\rightarrow $ $\overline{B}%
_{s}^{0}D_{s}^{-}$ is investigated by the same manner. Our results for the
strong coupling $g_{2}$ and partial width of this process read:%
\begin{equation}
g_{2}\equiv \mathcal{G}_{2}(-m_{D_{s}}^{2},m^{2})=(1.84\pm 0.32)\ \times
10^{-2}\ \mathrm{GeV}^{-1},
\end{equation}%
and
\begin{equation}
\Gamma \left[ \mathcal{M}_{\mathrm{b}}\rightarrow \overline{B}%
_{s}^{0}D_{s}^{-}\right] =(6.8\pm 1.8)~\mathrm{MeV}.
\end{equation}%
It is worth noting that the coupling $g_{2}$ has been found using the fit
function $\mathcal{G}_{2}(Q^{2},m^{2})$ with parameters $\mathcal{G}%
_{2}^{0}=0.02~\mathrm{GeV}^{-1}$, $c_{2}^{1}=4.74$, and $c_{2}^{2}=-6.42$.

The next channels of the hadronic molecule $\mathcal{M}_{\mathrm{b}}$ are
decays to the vector mesons' pairs $B^{\ast -}\overline{D}^{\ast 0}$, $%
\overline{B}^{\ast 0}D^{\ast -}$, $\overline{B}_{s}^{\ast 0}D_{s}^{\ast -}$.
As a sample, we analyze the mode $\mathcal{M}_{\mathrm{b}}\rightarrow
B^{\ast -}\overline{D}^{\ast 0}$ and write down formulas for this decay. The
correlator to be analyzed in this case is
\begin{eqnarray}
\Pi _{\mu \nu }(p,p^{\prime }) &=&i^{2}\int d^{4}xd^{4}ye^{ip^{\prime
}y}e^{-ipx}\langle 0|\mathcal{T}\{J_{\mu }^{B^{\ast }}(y)  \notag \\
&&\times J_{\nu }^{\overline{D}^{\ast }}(0)J^{\dagger }(x)\}|0\rangle .
\label{eq:CF3B}
\end{eqnarray}%
Here, $J_{\mu }^{B^{\ast }}(x)$ and $J_{\nu }^{\overline{D}^{\ast }}(x)$ are
currents which interpolate the vector particles $B^{\ast -}$ and $\overline{D%
}^{\ast 0}$%
\begin{equation}
J_{\mu }^{B^{\ast }}(x)=\overline{u}_{i}(x)\gamma _{\mu }b_{i}(x),\ J_{\nu
}^{\overline{D}^{\ast }}(x)=\overline{c}_{j}(x)\gamma _{\nu }u_{j}(x).
\end{equation}

To derive the physical side of the SR for the form factor $g_{3}(q^{2})$
describing the strong interactions of particles at the vertex $\mathcal{M}_{%
\mathrm{b}}B^{\ast -}\overline{D}^{\ast 0}$ we use the expression
\begin{eqnarray}
&&\Pi _{\mu \nu }^{\mathrm{Phys}}(p,p^{\prime })=\frac{\langle 0|J_{\mu
}^{B^{\ast }}|B^{\ast -}(p^{\prime },\varepsilon _{1})\rangle }{p^{\prime
2}-m_{B^{\ast }}^{2}}\frac{\langle 0|J_{\nu }^{\overline{D}^{\ast }}|%
\overline{D}^{\ast 0}(q,\varepsilon _{2})\rangle }{q^{2}-m_{D^{\ast }}^{2}}
\notag \\
&&\times \langle B^{\ast -}(p^{\prime },\varepsilon _{1})\overline{D}^{\ast
0}(q,\varepsilon _{2})|\mathcal{M}_{\mathrm{b}}(p)\rangle \frac{\langle
\mathcal{M}_{\mathrm{b}}(p)|J^{\dagger }|0\rangle }{p^{2}-m^{2}}  \notag \\
&&+\cdots .  \label{eq:CF3C}
\end{eqnarray}%
In Eq.\ (\ref{eq:CF3C}) $m_{B^{\ast }}$ and $m_{D^{\ast }}$ are the masses
of the final-state mesons, whereas $\varepsilon _{1}$ and $\varepsilon _{2}$
are their polarization vectors.

The correlation function $\Pi _{\mu \nu }^{\mathrm{Phys}}(p,p^{\prime })$
can be rewritten in the following form%
\begin{eqnarray}
&&\Pi _{\mu \nu }^{\mathrm{Phys}}(p,p^{\prime })=g_{3}(q^{2})\frac{\Lambda
f_{B^{\ast }}m_{B^{\ast }}f_{D^{\ast }}m_{D^{\ast }}}{\left(
p^{2}-m^{2}\right) (p^{\prime 2}-m_{B^{\ast }}^{2})}  \notag \\
&&\times \frac{1}{q^{2}-m_{D^{\ast }}^{2}}\left[ \frac{m^{2}-m_{B^{\ast
}}^{2}-q^{2}}{2}g_{\mu \nu }-p_{\nu }^{\prime }q_{\mu }\right]  \notag \\
&&+\cdots .
\end{eqnarray}%
This expression has been obtained by applying the matrix elements%
\begin{eqnarray}
&&\langle 0|J_{\mu }^{B^{\ast }}|B^{\ast -}(p^{\prime },\varepsilon
_{1})\rangle =f_{B^{\ast }}m_{B^{\ast }}\varepsilon _{1\mu },  \notag \\
&&\langle 0|J_{\nu }^{\overline{D}^{\ast }}|\overline{D}^{\ast
0}(q,\varepsilon _{2})\rangle =f_{D^{\ast }}m_{D^{\ast }}\varepsilon _{2\nu
},  \notag \\
&&\langle B^{\ast -}(p^{\prime },\varepsilon _{1})\overline{D}^{\ast
0}(q,\varepsilon _{2})|\mathcal{M}_{\mathrm{b}}(p)\rangle =g_{3}(q^{2})
\notag \\
&&\times \left[ q\cdot p^{\prime }\varepsilon _{1}^{\ast }\cdot \varepsilon
_{2}^{\ast }-q\cdot \varepsilon _{1}^{\ast }p^{\prime }\cdot \varepsilon
_{2}^{\ast }\right] .
\end{eqnarray}%
The QCD side of the SR is equal to
\begin{eqnarray}
&&\Pi _{\mu \nu }^{\mathrm{OPE}}(p,p^{\prime }) =\frac{1}{3}\int
d^{4}xd^{4}ye^{ip^{\prime }y}e^{-ipx}\langle \overline{b}b\rangle  \notag \\
&&\times \mathrm{Tr}\left[ \gamma _{\mu }S_{b}^{ia}(y-x)S_{c}^{aj}(x)\gamma
_{\nu }S_{u}^{ji}(-y)\right] .
\end{eqnarray}%
To find SR for the form factor $g_{3}(q^{2})$ we utilize amplitudes which
correspond to terms $\sim g_{\mu \nu }$ both in $\Pi _{\mu \nu }^{\mathrm{%
Phys}}(p,p^{\prime })$ and $\Pi _{\mu \nu }^{\mathrm{OPE}}(p,p^{\prime })$.
As a result, we get
\begin{eqnarray}
&&g_{3}(q^{2})=\frac{2(q^{2}-m_{D^{\ast }}^{2})}{\Lambda f_{B^{\ast
}}m_{B^{\ast }}f_{D^{\ast }}m_{D^{\ast }}(m^{2}-m_{B^{\ast }}^{2}-q^{2})}
\notag \\
&&\times e^{m^{2}/M_{1}^{2}}e^{m_{B^{\ast }}^{2}/M_{2}^{2}}\Pi _{3}(\mathbf{M%
}^{2},\mathbf{s}_{0},q^{2}),
\end{eqnarray}%
where $\Pi _{3}(\mathbf{M}^{2},\mathbf{s}_{0},q^{2})$ is transformed
amplitude $\Pi _{3}^{\mathrm{OPE}}(s,s^{\prime },q^{2})$ from $\Pi _{\mu \nu
}^{\mathrm{OPE}}(p,p^{\prime })$.

Numerical analysis is carried out by employing parameters of the particles $%
\mathcal{M}_{\mathrm{b}}$, $B^{\ast -}$, and $\overline{D}^{\ast 0}$ and
\begin{equation}
M_{2}^{2}\in \lbrack 5.5,6.5]~\mathrm{GeV}^{2},\ s_{0}^{\prime }\in \lbrack
34,35]~\mathrm{GeV}^{2}.
\end{equation}%
The parameters of the extrapolating function are $\mathcal{G}_{3}^{0}=0.022~%
\mathrm{GeV}^{-1}$, $c_{3}^{1}=10.65$, and $c_{3}^{2}=-19.06$. The coupling $%
g_{3}$ amounts to
\begin{equation}
g_{3}=(1.86\pm 0.35)\times 10^{-2}\ \mathrm{GeV}^{-1}.
\end{equation}%
Results obtained for $g_{3}(Q^{2})$ and fit function $\mathcal{G}%
_{3}(Q^{2},m^{2})$ are shown in Fig.\ \ref{fig:Fit1}.

We calculate the width of this decay by means of the formula%
\begin{equation}
\Gamma \left[ \mathcal{M}_{\mathrm{b}}\rightarrow B^{\ast -}\overline{D}%
^{\ast 0}\right] =g_{3}^{2}\frac{\lambda _{3}}{4\pi }\left( \lambda _{3}^{2}+%
\frac{3m_{B^{\ast }}^{2}m_{D^{\ast }}^{2}}{2m^{2}}\right) ,
\end{equation}%
where $\lambda _{3}=\lambda (m,m_{B^{\ast }},m_{D^{\ast }})$. This
expression leads to the prediction
\begin{equation}
\Gamma \left[ \mathcal{M}_{\mathrm{b}}\rightarrow B^{\ast -}\overline{D}%
^{\ast 0}\right] =(8.8\pm 2.4)~\mathrm{MeV}.
\end{equation}%
The widths of the decays $\mathcal{M}_{\mathrm{b}}\rightarrow \overline{B}%
^{\ast 0}D^{\ast -}$ and $\mathcal{M}_{\mathrm{b}}\rightarrow B^{\ast -}%
\overline{D}^{\ast 0}$ are equal to each other provided one neglects
differences in masses of the involved conventional mesons. Therefore, we
employ
\begin{equation}
\Gamma \left[ \mathcal{M}_{\mathrm{b}}\rightarrow \overline{B}^{0}D^{-}%
\right] \approx \Gamma \left[ \mathcal{M}_{\mathrm{b}}\rightarrow B^{-}%
\overline{D}^{0}\right] .
\end{equation}%
The process $\mathcal{M}_{\mathrm{b}}\rightarrow \overline{B}_{s}^{\ast
0}D_{s}^{\ast -}$ is studied by the similar manner. The coupling $g_{4}$ is
equal to
\begin{equation}
g_{4}=(1.72\pm 0.31)\times 10^{-2}\ \mathrm{GeV}^{-1},
\end{equation}%
extracted the parameters
\begin{equation}
M_{2}^{2}\in \lbrack 6,7]~\mathrm{GeV}^{2},\ s_{0}^{\prime }\in \lbrack
35,36]~\mathrm{GeV}^{2}.
\end{equation}%
Fot the partial width of this mode, we find%
\begin{equation}
\Gamma \left[ \mathcal{M}_{\mathrm{b}}\rightarrow \overline{B}_{s}^{\ast
0}D_{s}^{\ast -}\right] =(7.3\pm 1.9)~\mathrm{MeV}.
\end{equation}%
By taking into account all these decay channels, and results for their
partial widths it is not difficult to estimate the full decay width of the
hadronic molecule:
\begin{equation}
\Gamma \left[ \mathcal{M}_{\mathrm{b}}\right] =(93\pm 17)~\mathrm{MeV}.
\end{equation}

It has been emphasized above that at lower mass ($\mathrm{l.m.}$) $m=15638~%
\mathrm{MeV}$ the molecule $\mathcal{M}_{\mathrm{b}}$ is stable against the
dominant decay channel. As a result, its full decay width is formed due to
subleading processes. To evaluate $\Gamma \left[ \mathcal{M}_{\mathrm{b}}%
\right] |_{\mathrm{l.m.}}$, we have repeated calculations of the current
subsection with $m=15638~\mathrm{MeV}$. Our prediction for the width of the
molecule $\mathcal{M}_{\mathrm{b}}$ reads
\begin{equation}
\Gamma \left[ \mathcal{M}_{\mathrm{b}}\right] |_{\mathrm{l.m.}}=(49\pm 6)~%
\mathrm{MeV}.
\end{equation}

%%%%%%%%%%%%%%%%%%%%%%%%%%%%%%%%%%%%%%%%%%%%%%%%%%%%%%%%%%%%%%%%%

\section{Width of the molecule $\mathcal{M}_{\mathrm{c}}$}

\label{sec:Widths2}

%%%%%%%%%%%%%%%%%%%%%%%%%%%%%%%%%%%%%%%%%%%%%%%%%%%%%%%%%%%

Here, we evaluate the width of the molecule $\mathcal{M}_{\mathrm{c}}$ by
studying its dominant modes $\mathcal{M}_{\mathrm{c}}\rightarrow \eta
_{c}B_{c}^{+}$ and $\mathcal{M}_{\mathrm{c}}\rightarrow J/\psi B_{c}^{\ast
+} $, as well as six channels generated by $c\overline{c}$ annihilation in $%
\mathcal{M}_{\mathrm{c}}$. It is clear that the both dominant decays are
permitted channels for $\mathcal{M}_{\mathrm{c}}$. Indeed, the mass $%
\widetilde{m}=9712~\mathrm{MeV}$ of $\mathcal{M}_{\mathrm{c}}$ exceeds
thresholds for these processes which amount to $9259~\mathrm{MeV}$ and $9435~%
\mathrm{MeV}$. Even in the lower limit $\widetilde{m}=9640~\mathrm{MeV}$
these decays are allowed channels of $\mathcal{M}_{\mathrm{c}}$. In this
sense, it differs from the hadronic molecule $\mathcal{M}_{\mathrm{b}}$.

Investigation of the decay $\mathcal{M}_{\mathrm{c}}\rightarrow \eta
_{c}B_{c}^{+}$ does not differ considerably from analysis performed in the
previous section. Here, we should calculate the form factor $\widetilde{g}%
_{1}(q^{2})$ and find the strong coupling $\widetilde{g}_{1}$ at the vertex $%
\mathcal{M}_{\mathrm{c}}\eta _{c}B_{c}^{+}$. We start to consider the
correlator
\begin{eqnarray}
\widetilde{\Pi }(p,p^{\prime }) &=&i^{2}\int d^{4}xd^{4}ye^{ip^{\prime
}y}e^{-ipx}\langle 0|\mathcal{T}\{J^{B_{c}^{+}}(y)  \notag \\
&&\times J^{\eta _{c}}(0)\widetilde{J}^{\dagger }(x)\}|0\rangle ,
\end{eqnarray}%
with $J^{B_{c}^{+}}(x)$ and $J^{\eta _{c}}(x)$ being the interpolating
currents for the mesons $B_{c}^{+}$ and $\eta _{c}$, respectively%
\begin{equation}
\ J^{B_{c}^{+}}(x)=\overline{b}_{i}(x)i\gamma _{5}c_{i}(x),\ J^{\eta
_{c}}(x)=\overline{c}_{j}(x)i\gamma _{5}c_{j}(x).
\end{equation}%
We compute the physical side of SR using the following matrix elements%
\begin{eqnarray}
\langle 0|J^{\eta _{c}}|\eta _{c}(q)\rangle &=&\frac{f_{\eta _{c}}m_{\eta
_{c}}^{2}}{2m_{c}},  \notag \\
\langle 0|J^{B_{c}^{+}}|B_{c}^{+}(p^{\prime })\rangle &=&\frac{%
f_{B_{c}}m_{B_{c}}^{2}}{m_{b}+m_{c}},
\end{eqnarray}%
and
\begin{equation}
\langle \eta _{c}(q)B_{c}^{+}(p^{\prime })|\mathcal{M}_{\mathrm{c}%
}(p)\rangle =\widetilde{g}_{1}(q^{2})p\cdot p^{\prime }.
\end{equation}%
In the formulas above, the mass and decay constant of the pseudoscalar meson
$\eta _{c}$ are denoted as $m_{\eta _{c}}$ and $f_{\eta _{c}}$, respectively.

The phenomenological and QCD components of this SR have analytical forms
presented in Sec.\ \ref{sec:Widths1} with evident replacements. As a result,
the SR for $\widetilde{g}_{1}(q^{2})$ reads%
\begin{eqnarray}
\widetilde{g}_{1}(q^{2}) &=&\frac{4m_{c}(m_{b}+m_{c})(q^{2}-m_{\eta
_{c}}^{2})}{\Lambda f_{\eta _{c}}m_{\eta
_{c}}^{2}f_{B_{c}}m_{B_{c}}^{2}(m^{2}+m_{B_{c}}^{2}-q^{2})}  \notag \\
&&\times e^{m^{2}/M_{1}^{2}}e^{m_{B_{c}}^{2}/M_{2}^{2}}\widetilde{\Pi }_{1}(%
\mathbf{M}^{2},\mathbf{s}_{0},q^{2}).  \label{eq:SRg1}
\end{eqnarray}%
The correlation function $\widetilde{\Pi }_{1}(\mathbf{M}^{2},\mathbf{s}%
_{0},q^{2})$ is determined by the expression
\begin{eqnarray}
\widetilde{\Pi }(\mathbf{M}^{2},\mathbf{s}_{0},q^{2})
&=&\int_{(m_{b}+3m_{c})^{2}}^{s_{0}}\int_{(m_{b}+m_{c})^{2}}^{s_{0}^{\prime
}}dsds^{\prime }e^{-s/M_{1}^{2}}  \notag \\
&&\times e^{-s^{\prime }/M_{2}^{2}}\widetilde{\rho }(s,s^{\prime },q^{2}).
\end{eqnarray}

In calculations the parameters $(M_{1}^{2},s_{0})$ in the channel of the
molecule $\mathcal{M}_{\mathrm{c}}$ are chosen as in Eq.\ (\ref{eq:Wind1A}).
The intervals for $(M_{2}^{2},s_{0}^{\prime })$ in the $B_{c}^{+}$ channel
are%
\begin{equation}
M_{2}^{2}\in \lbrack 6.5,7.5]~\mathrm{GeV}^{2},\ s_{0}^{\prime }\in \lbrack
45,47]~\mathrm{GeV}^{2}.  \label{eq:Wind4}
\end{equation}%
The function $\widetilde{g}_{1}(Q^{2})$ is calculated at $Q^{2}=2-20~\mathrm{%
GeV}^{2}$. The extrapolating function $\widetilde{\mathcal{G}}_{1}(Q^{2},%
\widetilde{m}^{2})$ has the form Eq.\ (\ref{eq:FitF}) with $m^{2}$
substituted by $\widetilde{m}^{2}$. The function $\widetilde{\mathcal{G}}%
_{1} $ has the parameters $\widetilde{\mathcal{G}}_{1}^{0}=0.132~\mathrm{GeV}%
^{-1},\widetilde{c}_{1}^{1}=3.148,$ and $\widetilde{c}_{1}^{2}=-2.152$.

The coupling $\widetilde{g}_{1}$ extracted at the mass shell $q^{2}=m_{\eta
_{c}}^{2}$ amounts to
\begin{equation}
\widetilde{g}_{1}\equiv \widetilde{\mathcal{G}}_{1}(-m_{\eta _{c}}^{2},%
\widetilde{m}^{2})=(9.63\pm 1.86)\times 10^{-2}\ \mathrm{GeV}^{-1}.
\end{equation}%
We evaluate the partial width of this channel by employing the expression
\begin{equation}
\Gamma \left[ \mathcal{M}_{\mathrm{c}}\rightarrow \eta _{c}B_{c}^{+}\right] =%
\widetilde{g}_{1}^{2}\frac{m_{B_{c}}^{2}\widetilde{\lambda }_{1}}{8\pi }%
\left( 1+\frac{\widetilde{\lambda }_{1}^{2}}{m_{B_{c}}^{2}}\right) ,
\end{equation}%
where $\widetilde{\lambda }_{1}$ is $\lambda (\widetilde{m}%
,m_{B_{c}},m_{\eta _{c}})$. Our prediction is%
\begin{equation}
\Gamma \left[ \mathcal{M}_{\mathrm{c}}\rightarrow \eta _{c}B_{c}^{+}\right]
=(21.0\pm 6.0)~\mathrm{MeV}.  \label{eq:DW3}
\end{equation}

The second dominant channel of the molecule $\mathcal{M}_{\mathrm{c}}$ is
the decay to particles $J/\psi $ and $B_{c}^{\ast +}$. To find the coupling $%
\widetilde{g}_{2}$ at the vertex $\mathcal{M}_{\mathrm{c}}J/\psi B_{c}^{\ast
+}$, one should compute the relevant form factor $\widetilde{g}_{2}(q^{2})$,
which can obtained from the sum rule for this function. To this end, we
consider the correlator
\begin{eqnarray}
\widetilde{\Pi }_{\mu \nu }(p,p^{\prime }) &=&i^{2}\int
d^{4}xd^{4}ye^{ip^{\prime }y}e^{-ipx}\langle 0|\mathcal{T}\{J_{\mu
}^{B_{c}^{\ast }}(y)  \notag \\
&&\times J_{\nu }^{J/\psi }(0)\widetilde{J}^{\dagger }(x)\}|0\rangle ,
\label{eq:CF4}
\end{eqnarray}%
with $J_{\mu }^{B_{c}^{\ast }}(x)$ and $J_{\nu }^{J/\psi }(x)$ being the
currents that interpolate vector mesons $B_{c}^{\ast +}$ and $J/\psi $,
respectively
\begin{equation}
J_{\mu }^{B_{c}^{\ast }}(x)=\overline{b}_{i}(x)\gamma _{\mu }c_{i}(x),\
J_{\nu }^{J/\psi }(x)=\overline{c}_{j}(x)\gamma _{\nu }c_{j}(x).
\end{equation}

The phenomenological side of SR $\Pi _{\mu \nu }^{\mathrm{Phys}}(p,p^{\prime
})$ is given by the standard expression%
\begin{eqnarray}
&&\Pi _{\mu \nu }^{\mathrm{Phys}}(p,p^{\prime })=\frac{\langle 0|J_{\mu
}^{B_{c}^{\ast }}|B_{c}^{\ast +}(p^{\prime },\epsilon _{1})\rangle }{%
p^{\prime 2}-m_{B_{c}^{\ast }}^{2}}\frac{\langle 0|J_{\nu }^{J/\psi }|J/\psi
(q,\epsilon _{2})\rangle }{q^{2}-m_{J/\psi }^{2}}  \notag \\
&&\times \langle B_{c}^{\ast +}(p^{\prime },\epsilon _{1})J/\psi (q,\epsilon
_{2})|\mathcal{M}_{\mathrm{c}}(p)\rangle \frac{\langle \mathcal{M}_{\mathrm{c%
}}(p)|\widetilde{J}^{\dagger }|0\rangle }{p^{2}-\widetilde{m}^{2}}+\cdots .
\notag \\
&&  \label{eq:CF5}
\end{eqnarray}%
Here, $m_{J/\psi }$ and $m_{B_{c}^{\ast }}$ are the masses of the mesons,
and $\epsilon _{1},$ $,\epsilon _{2}$- the polarization vectors of these
particles.

The correlator $\Pi _{\mu \nu }^{\mathrm{Phys}}$ can be rewritten by using
the matrix elements
\begin{eqnarray}
\langle 0|J_{\mu }^{B_{c}^{\ast }}|B_{c}^{\ast +}(p^{\prime })\rangle
&=&f_{B_{c}^{\ast }}m_{B_{c}^{\ast }}\epsilon _{1\mu },  \notag \\
\langle 0|J_{\nu }^{J/\psi }|J/\psi (q)\rangle &=&f_{J/\psi }m_{J/\psi
}\epsilon _{2\nu },  \label{eq:ME2}
\end{eqnarray}%
and
\begin{eqnarray}
&&\langle B_{c}^{\ast +}(p^{\prime },\epsilon _{1})J/\psi (q,\epsilon _{2})|%
\mathcal{M}_{\mathrm{c}}(p)\rangle =\widetilde{g}_{2}(q^{2})  \notag \\
&&\times \left[ q\cdot p^{\prime }\epsilon _{1}^{\ast }\cdot \epsilon
_{2}^{\ast }-q\cdot \epsilon _{1}^{\ast }p^{\prime }\cdot \epsilon
_{2}^{\ast }\right] .  \label{eq:ME1B}
\end{eqnarray}%
In Eq.\ (\ref{eq:ME1B}) $f_{J/\psi }$ and $f_{B_{c}^{\ast }}$ are the decay
constants of $J/\psi $ and $B_{c}^{\ast +}$, respectively.

Then, for $\widetilde{\Pi }_{\mu \nu }^{\mathrm{Phys}}(p,p^{\prime })$ we
get
\begin{eqnarray}
&&\widetilde{\Pi }_{\mu \nu }^{\mathrm{Phys}}(p,p^{\prime })=\frac{%
\widetilde{g}_{2}(q^{2})\widetilde{\Lambda }f_{B_{c}^{\ast }}m_{B_{c}^{\ast
}}f_{J/\psi }m_{J/\psi }}{(p^{2}-\widetilde{m}^{2})(p^{\prime
2}-m_{B_{c}^{\ast }}^{2})(q^{2}-m_{J/\psi }^{2})}  \notag \\
&&\times \left[ \frac{(m^{2}-m_{B_{c}^{\ast }}^{2}-q^{2})}{2}g_{\mu \nu
}-q_{\mu }p_{\nu }^{\prime }+\cdots \right] .  \notag \\
&&  \label{eq:CF5a}
\end{eqnarray}%
The correlation function $\widetilde{\Pi }_{\mu \nu }(p,p^{\prime })$
expressed using quark propagators becomes equal to
\begin{eqnarray}
&&\widetilde{\Pi }_{\mu \nu }^{\mathrm{OPE}}(p,p^{\prime })=i^{2}\int
d^{4}xd^{4}ye^{ip^{\prime }y}e^{-ipx}\mathrm{Tr}\left[ \gamma _{\mu
}S_{c}^{ja}(-x)\right.  \notag \\
&&\left. \times \gamma _{5}S_{c}^{ai}(x-y)\gamma _{\nu
}S_{c}^{ib}(y-x)\gamma _{5}S_{b}^{bj}(x)\right] .  \label{eq:CF6}
\end{eqnarray}%
To derive SR for the form factor $\widetilde{g}_{2}(q^{2})$, we utilize the
invariant amplitudes which correspond to terms proportional to $g_{\mu \nu }$
in Eqs.\ (\ref{eq:CF5a}) and (\ref{eq:CF6}). Then, we find for $\widetilde{g}%
_{2}(q^{2})$%
\begin{eqnarray}
\widetilde{g}_{2}(q^{2}) &=&\frac{2(q^{2}-m_{J/\psi }^{2})}{\widetilde{%
\Lambda }f_{B_{c}^{\ast }}m_{B_{c}^{\ast }}f_{J/\psi }m_{J/\psi
}(m^{2}-m_{B_{c}^{\ast }}^{2}-q^{2})}  \notag \\
&&\times e^{m^{2}/M_{1}^{2}}e^{m_{B_{c}^{\ast }}^{2}/M_{2}^{2}}\widetilde{%
\Pi }_{2}(\mathbf{M}^{2},\mathbf{s}_{0},q^{2}).
\end{eqnarray}

Operations to find the coupling $\widetilde{g}_{2}$ have been explained
above so we give final results without details. Note that the function $%
\widetilde{g}_{2}(Q^{2})$ is calculated for $Q^{2}=2-30~\mathrm{GeV}^{2}$.
In the $\mathcal{M}_{\mathrm{c}}$ channel parameters $(M_{1}^{2},s_{0})$ are
chosen as in Eq.\ (\ref{eq:Wind1A}). In the $B_{c}^{\ast +}$ channel, we
have varied $(M_{2}^{2},s_{0}^{\prime })$ inside windows
\begin{equation}
M_{2}^{2}\in \lbrack 6.5,7.5]~\mathrm{GeV}^{2},\ s_{0}^{\prime }\in \lbrack
50,51]~\mathrm{GeV}^{2}.
\end{equation}%
The function $\widetilde{\mathcal{G}}_{2}(Q^{2},\widetilde{m}^{2})$ is fixed
by constants: $\widetilde{\mathcal{G}}_{2}^{0}=0.40~\mathrm{GeV}^{-1},%
\widetilde{c}_{2}^{1}=6.88,$and $\widetilde{c}_{2}^{2}=-5.65$. Then, the
coupling $\widetilde{g}_{2}$ is equal to
\begin{equation}
\widetilde{g}_{2}\equiv \widetilde{\mathcal{G}}_{2}(-m_{J/\psi }^{2},%
\widetilde{m}^{2})=(1.9\pm 0.4)\times 10^{-1}\ \mathrm{GeV}^{-1}.
\end{equation}

The width of the decay $\mathcal{M}_{\mathrm{c}}\rightarrow J/\psi
B_{c}^{\ast +}$ is obtained using the formula
\begin{equation}
\Gamma \left[ \mathcal{M}_{\mathrm{c}}\rightarrow J/\psi B_{c}^{\ast +}%
\right] = \widetilde{g}_{2}^{2}\frac{ \widetilde{\lambda} _{2}}{4\pi }\left(  \widetilde{\lambda} _{2}^{2}+\frac{%
3m_{B_{c}^{\ast }}^{2}m_{J/\psi }^{2}}{2\widetilde{m}^{2}}\right) ,
\end{equation}%
where $ \widetilde{\lambda} _{2}$ is $\lambda (\widetilde{m},m_{B_{c}^{\ast }},m_{J/\psi
})$. We find%
\begin{equation}
\Gamma \left[ \mathcal{M}_{\mathrm{c}}\rightarrow J/\psi B_{c}^{\ast +}%
\right] =(22.1\pm 7.5)~\mathrm{MeV}.
\end{equation}
\

We have explored also six decay channels $\mathcal{M}_{\mathrm{c}%
}\rightarrow B^{+}D^{0}$, $B^{0}D^{+}$, $B_{s}^{0}D_{s}^{+}$, $B^{\ast
+}D^{\ast 0}$, $B^{\ast 0}D^{\ast +}$, and $B_{s}^{\ast 0}D_{s}^{\ast +}$
triggered by annihilation of $c\overline{c}$ quarks. We have benefited from
the facts $\Gamma \left[ \mathcal{M}_{\mathrm{c}}\rightarrow B^{+}D^{0}%
\right] \approx \Gamma \left[ \mathcal{M}_{\mathrm{c}}\rightarrow B^{0}D^{+}%
\right] $ and $\Gamma \left[ \mathcal{M}_{\mathrm{c}}\rightarrow B^{\ast
+}D^{\ast 0}\right] \approx \Gamma \left[ \mathcal{M}_{\mathrm{c}%
}\rightarrow B^{\ast 0}D^{\ast +}\right] $. Final information on remaining
four channels are presented in Table\ \ref{tab:Channels}.

The full width of the molecule $\mathcal{M}_{\mathrm{c}}$ saturated by these
decay channels is%
\begin{equation}
\Gamma \left[ \mathcal{M}_{\mathrm{c}}\right] =(70\pm 10)~\mathrm{MeV}.
\end{equation}

\begin{table}[tbp]
\begin{tabular}{|c|c|c|c|}
\hline\hline
i & Channels & $\widetilde{g}_{i}\times 10^{2}~(\mathrm{GeV}^{-1})$ & $%
\Gamma_{i}~(\mathrm{MeV})$ \\ \hline
$1$ & $B^{+}D^{0}$ & $3.2 \pm 0.6$ & $4.8 \pm 1.3$ \\
$2$ & $B^{0}_{s}D^{+}_{s}$ & $2.9 \pm 0.5$ & $3.7 \pm 0.9$ \\
$3$ & $B^{\ast +}D^{\ast 0}$ & $4.3 \pm 0.7$ & $4.8 \pm 1.2 $ \\
$4$ & $B_{s}^{\ast 0}D_{s}^{\ast +}$ & $4.1 \pm 0.6$ & $4.0 \pm 0.9 $ \\
\hline\hline
\end{tabular}%
\caption{Decay channels of the molecule $\mathcal{M}_{c}$ due to $c\overline{%
c}$ annihilation, corresponding strong couplings $\widetilde{g}_{i}$ and
widths $\Gamma _{i}$.}
\label{tab:Channels}
\end{table}

%%%%%%%%%%%%%%%%%%%%%%%%%%%%%%%%%%%%%%%%%%%%%%%%%%%%%%%%%%%%%%

\section{Conclusions}

\label{sec:Conc}

%%%%%%%%%%%%%%%%%%%%%%%%%%%%%%%%%%%%%%%%%%%%%%%%%%%%%%%%%%%

Investigations carried out in the present work is a new step towards
understanding of the internal structure and properties of the potential all
heavy four-quark mesons. We have considered the scalar structures $bb%
\overline{b}\overline{c}$ and $cc\overline{c}\overline{b}$ organized as
hadronic molecules $\mathcal{M}_{\mathrm{b}}=\eta _{b}B_{c}^{-}$ and $%
\mathcal{M}_{\mathrm{c}}=\eta _{c}B_{c}^{+}$. We have calculated their
masses and evaluated decay widths by analyzing the dominant and subleading
decay channels.

The masses and current couplings of these molecules have been calculated by
means of QCD two-point sum rule method. The central values of the
predictions $m=(15728\pm 90)~\mathrm{MeV}$ and $\widetilde{m}=(9712\pm 72)~%
\mathrm{MeV}$ mean that $\mathcal{M}_{\mathrm{b}}$ is close to $\eta
_{b}B_{c}^{-}$ threshold, whereas $\mathcal{M}_{\mathrm{c}}$ locates a few
hundred $\mathrm{MeV}$ above corresponding borders. In the lower limit for $%
m=15638~\mathrm{MeV}$ the structure $\mathcal{M}_{\mathrm{b}}$ can be
interpreted as a bound state of the mesons $\eta _{b}$ and $B_{c}^{-}$.
Contrary, $\mathcal{M}_{\mathrm{c}}$ does not form a bound state and is a
broad resonance above relevant two-meson continuum.

Predictions for the masses of $\mathcal{M}_{\mathrm{b}}$ and $\mathcal{M}_{%
\mathrm{c}}$ have allowed us to determine their possible decay channels. In
our studies we have distinguished the dominant and subleading decay
mechanisms of these particles. The dominant mechanism is one in which all
constituent quarks participate in producing of ordinary final-state mesons.
For molecule $\mathcal{M}_{\mathrm{b}}$ breakdown to $\eta _{b}$ and $%
B_{c}^{-}$ mesons is the dominant process. The dominant channels of $%
\mathcal{M}_{\mathrm{c}}$ are the processes $\mathcal{M}_{\mathrm{c}%
}\rightarrow \eta _{c}B_{c}^{+}$ and $\mathcal{M}_{\mathrm{c}}\rightarrow
J/\psi B_{c}^{\ast +}$. In the last decay $\mathcal{M}_{\mathrm{c}}$ falls
to vector partners of the constituent mesons.

Another mechanism of decays is generated by annihilation of constituent $b%
\overline{b}$ or $c\overline{c}$ quarks inside of the molecules $\mathcal{M}%
_{\mathrm{b}}$ and $\mathcal{M}_{\mathrm{c}}$ and producing $B_{(s)}^{(\ast
)}D_{(s)}^{(\ast )}$ pairs with appropriate charges and spin-parities. This
mechanism has been included into the SR framework after replacing in the
correlation functions the vacuum expectation values $m_{b}\langle \overline{b%
}b\rangle $ and $m_{c}\langle \overline{c}c\rangle $ by a term $\sim \langle
\alpha _{s}G^{2}/\pi \rangle $. It is worth emphasizing that relations
between the heavy quark and gluon condensates were extracted within the SR
method and are approximate expressions.

All decay channels considered in this work have been explored using the
three-point SR approach. It has permitted us to estimate the strong
couplings $g_{i}$ and $\widetilde{g}_{i}$ at the vertices $\mathcal{M}_{%
\mathrm{b}}M_{1}M_{2}$ and $\mathcal{M}_{\mathrm{c}}M_{1}M_{2}$, where $%
M_{1} $ and $M_{2}$ are the final-state mesons. Our predictions $\Gamma %
\left[ \mathcal{M}_{\mathrm{b}}\right] =(93\pm 17)~\mathrm{MeV}$ and $\Gamma %
\left[ \mathcal{M}_{\mathrm{c}}\right] =(70\pm 10)~\mathrm{MeV}$ for the
widths of the molecules $\mathcal{M}_{\mathrm{b}}$ and $\mathcal{M}_{\mathrm{%
c}}$ imply that they may be interpreted as relatively broad structures. Note
that numerous subleading processes form sizeable parts of these parameters.
In the scenario when $m=15638~\mathrm{MeV}$ the mesons $\eta _{b}$ and $%
B_{c}^{-}$ form the bound state $\mathcal{M}_{\mathrm{b}}$ which is,
nevertheless, unstable hadronic molecule with $\Gamma \left[ \mathcal{M}_{%
\mathrm{b}}\right] |_{\mathrm{l.m.}}=(49\pm 6)~\mathrm{MeV}.$

As it has been emphasized in Sec.\ \ref{sec:Intro} the exotic scalar mesons $%
T_{\mathrm{b}}$ and $T_{\mathrm{c}}$ with the same contents but
diquark-antidiquark structures were explored in our work \cite{Agaev:2024uza}%
. It is interesting to compare parameters of these states with ones obtained
in the present article. Thus, the tetraquarks $T_{\mathrm{b}}$ and $T_{%
\mathrm{c}}$ have the masses $(15698\pm 95)~\mathrm{MeV}$ and $(9680\pm 102)~%
\mathrm{MeV}$, respectively. In other words, the molecules $\mathcal{M}_{%
\mathrm{b}}$ and $\mathcal{M}_{\mathrm{c}}$ are heavier than their
diquark-antidiquark counterparts. This is connected with the internal
organization of the molecule and diquark-antidiuark structures. A hadronic
molecule is composed of the color-neutral mesons, while in the diquark
picture four-quark meson is formed owing to interaction of colored diquark
and antidiquark which establish tightly bound state. The particles $T_{%
\mathrm{b}}$ and $T_{\mathrm{c}}$ with widths $\Gamma \lbrack T_{\mathrm{b}%
}]=(36.0\pm 10.4)~\mathrm{MeV}$ and $\Gamma \lbrack T_{\mathrm{c}}]=(54.7\pm
12.6)~\mathrm{MeV}$ are narrower than molecules $\mathcal{M}_{\mathrm{b}}$
and $\mathcal{M}_{\mathrm{c}}$ because of the same reason. But here one
should take into account that widths of the tetraquarks $T_{\mathrm{b}}$ and
$T_{\mathrm{c}}$  were estimated by analyzing their dominant decay channels.
In the lower mass limit $15603~\mathrm{MeV}$ the tetraquark $T_{\mathrm{b}}$
is also stable against decay to $\eta _{b}$ and $B_{c}^{-}$ mesons. Its
width in this case is formed due to subleading processes: This problem may
be addressed in our future publication(s).

The hadronic molecules composed of four $b$ and $c$ quarks in various
combinations were studied in Ref.\ \cite{Liu:2024pio}. There, authors used
the local gauge formalism to investigate the meson-meson interactions in
such systems. In the scalar sector of this model, the molecular states rest
above the relevant two-meson thresholds. Our findings for the scalar
molecules $\mathcal{M}_{\mathrm{b}}$ and $\mathcal{M}_{\mathrm{c}}$ are in
part consistent with this conclusion of Ref.\ \cite{Liu:2024pio}. In this
article the authors gave also information on parameter-dependent masses of
axial-vector molecules $\Upsilon B_{c}^{-}$, $\eta _{b}B_{c}^{\ast -}$, and $%
\Upsilon B_{c}^{\ast -}$ which lie below the corresponding two-meson
thresholds. It other words, these molecules can not dissociate to their
ingredients, and in this sense, are stable structures. Of course, this does
not mean that they are stable against the strong decays through annihilation
mechanisms, which may lead to considerably broad structures even in these
cases. Predictions of Ref.\ \cite{Liu:2024pio} are interesting for
understanding the internal organizations and binding mechanisms of the fully
heavy hadronic molecules, but need to be confirmed using alternative
approaches including the sum rule method. This problem is beyond the scope
of the present article, but eventually may be addressed in our future works.

The studies carried out in the present paper provide valuable information on
parameters of hadronic molecules built of heavy quarks and may be useful for
experimental analysis of such systems.

\end{document}